\documentclass[a4paper,aps,amsmath,amssymb,twocolumn,longbibliography,,accepted=2022-05-17]{quantumarticle}
\pdfoutput=1
\usepackage[utf8]{inputenc}
\usepackage[english]{babel}
\usepackage[T1]{fontenc}
\usepackage{amsmath}
\usepackage{hyperref}

\usepackage{tikz}
\usepackage{lipsum}

\begin{document}


\title{Deep Learning of Quantum Many-Body Dynamics via \newline Random Driving}
\author{Naeimeh Mohseni}
\affiliation {Max-Planck-Institut f{\"u}r die Physik des Lichts, Staudtstrasse 2, 91058 Erlangen, Germany}
\affiliation{Physics Department, University of Erlangen-Nuremberg, Staudtstr. 5, 91058 Erlangen, Germany}
\author{Thomas Fösel}
\affiliation {Max-Planck-Institut f{\"u}r die Physik des Lichts, Staudtstrasse 2, 91058 Erlangen, Germany}
\affiliation{Physics Department, University of Erlangen-Nuremberg, Staudtstr. 5, 91058 Erlangen, Germany}
\author{Lingzhen Guo}
\affiliation {Max-Planck-Institut f{\"u}r die Physik des Lichts, Staudtstrasse 2, 91058 Erlangen, Germany}
\author{Carlos Navarrete-Benlloch}
\altaffiliation{Corresponding author: derekkorg@gmail.com}
\affiliation{Wilczek Quantum Center, School of Physics and Astronomy, Shanghai Jiao Tong University, Shanghai 200240, China}
\affiliation{Shanghai Research Center for Quantum Sciences, Shanghai 201315, China}
\affiliation {Max-Planck-Institut f{\"u}r die Physik des Lichts, Staudtstrasse 2, 91058 Erlangen, Germany}
\author{Florian Marquardt}
\affiliation {Max-Planck-Institut f{\"u}r die Physik des Lichts, Staudtstrasse 2, 91058 Erlangen, Germany}
\affiliation{Physics Department, University of Erlangen-Nuremberg, Staudtstr. 5, 91058 Erlangen, Germany}
\maketitle
\begin{abstract}
    Neural networks have emerged as a powerful way to approach many practical problems in quantum physics. In this work, we illustrate the power of deep learning to predict the dynamics of a quantum many-body system, where the training is \textit{based purely on monitoring expectation values of observables under random driving}. The trained recurrent network is able to produce accurate predictions for driving trajectories entirely different than those observed during training. As a proof of principle, here we train the network on numerical data generated from spin models, showing that it can learn the dynamics of observables of interest without needing information about the full quantum state. This allows our approach to be applied eventually to actual experimental data generated from a quantum many-body system that might be open, noisy, or disordered, without any need for a detailed understanding of the system. This scheme provides considerable speedup for rapid explorations and pulse optimization. Remarkably, we show the network is able to extrapolate the dynamics to times longer than those it has been trained on, as well as to the infinite-system-size limit.
\end{abstract}
\section{Introduction\label{sec:introduction}}
Machine learning based on neural networks (NN) \cite{lecun2015deep,goodfellow2016deep} has revealed a remarkable potential in solving a wide variety of complex problems. More recently, a number of first applications to quantum physics have emerged. These include the identification of quantum phases of matter and learning phase transitions \cite{PhysRevB.94.195105, carrasquilla2017machine,van2017learning, wetzel2017unsupervised, beach2018machine}, quantum state tomography \cite{torlai2018neural,Prx}, quantum error correction and decoding \cite{PhysRevLett.119.030501,PhysRevX.8.031084}, improving quantum Monte Carlo methods \cite{PhysRevResearch.2.012039}, solving optimization problems encoded in the ground state of a quantum many-body Hamiltonian \cite{ mills2020controlled}, and tackling quantum many-body dynamics \cite{carleo2017solving,gao2017efficient, schmitt2019quantum, mohseni2021deep}. Our work provides a step forward in the latter direction.

\begin{figure}[h!]
	\centering
	\includegraphics[width=1.0\linewidth]{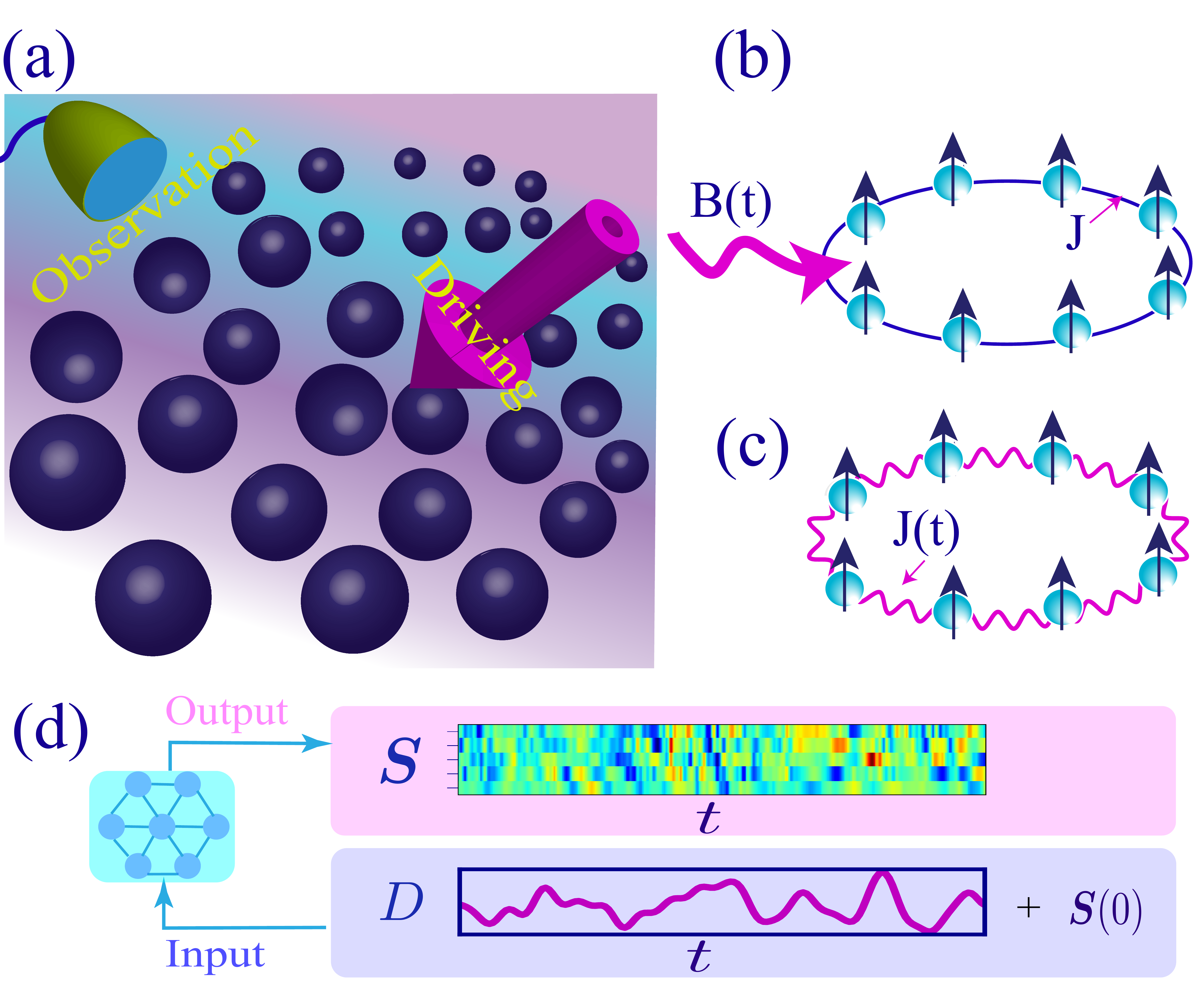}
	\caption{Scenario considered in this work.
		(a) Schematic representation of a quantum many-body system subject to an arbitrary driving field, where the goal is to predict the time-evolution of a set of observables. The models that we explore here, as for illustration, are spin ring subject to
		(b) time-dependent magnetic field or
		(c) time-dependent exchange couplings.
		(d) Schematic representation of input and output of the NN. The input of the NN is the driving trajectory $D(t)$, together with other time-independent parameters and a description of the initial state of the system shown by  $\boldsymbol{S}_0$. As output the NN predicts the full evolution of expectation values of observables of interest, generically denoted by $\boldsymbol{S}(t)$ in the figure. This includes local observables but may also comprise, e.g., spatial two-point correlators.
	}
	\label{fig1}
\end{figure}

In recent years, there has been tremendous activity in the field of non-equilibrium engineered quantum many-body systems, particularly motivated by the significant progress achieved in experiments \cite{richerme2014non, greiner2002collapse, PhysRevLett.111.053003, kinoshita2006quantum,martinez2016real, richerme2014non,cheneau2012light}. However, simulating the dynamics of quantum many-body systems computationally is extremely difficult, even in 1D for long times. The reason is that, besides the exponential growth of the Hilbert space dimension with the system size, quantum correlations also grow significantly with time \cite{Calabrese_2005}. Recently, it has been shown that machine learning based on NNs can assist in overcoming such difficulties to a great extent \cite{carleo2017solving, schmitt2019quantum, lopez2019real}. For example, for systems that can be described by local Hamiltonians, dynamics is confined in a smaller sector of the Hilbert space, so that a majority of quantum states
cannot be reached during the time interval of interest \cite{PhysRevLett.106.170501}. In such cases, it might be enough to find an appropriate variational ansatz for the state of system, containing a number of time-dependent parameters that may scale polynomially with the system  size. NNs have shown great potential as such variational ansatzes for the quantum state \cite{carleo2017solving, gao2017efficient, schmitt2019quantum}. Nevertheless, challenges remain with this approach, e.g. in efficiently evaluating expectation values.

In this work, we introduce a different approach to exploiting NNs for describing driven quantum many-body dynamics.
Our aim is to teach a NN to predict the time-evolution of observables in non-equilibrium quantum many-body systems subject to arbitrary driving. Such driven many-body models play a key role in many areas, such as cold atom experiments \cite{struck2013engineering,RevModPhys.89.011004}, NMR \cite{foroozandeh2014ultrahigh,RevModPhys.76.1037}, and quantum information processing \cite{anderlini2007controlled}.

In our approach, the learning is \textit{based purely on observations of expectation values during randomly driven dynamics} (Fig.~\ref{fig1}). If successful, this means such NNs can eventually be trained on actual experimental systems, \textit{without any prior knowledge} of the underlying Hamiltonian or of any additional dissipative couplings and other aspects of the real physical system. It also means that they can be trained on \textit{partial} observations, i.e., measurements of only a selected subset of degrees of freedom.


It is obvious that this goal represents a considerable challenge: Even if the Hamiltonian were known, predicting the full time evolution of a quantum many-body system in principle requires knowledge of the \textit{exponentially large} many-body wave function. Alternatively, one has to come up with an efficient compressed representation, with matrix product states \cite{orus2014practical} as one prominent typical example. It is, however, hard to find such compressed representations in general, and their applicability can depend very much on circumstances such as the space dimension. In addition, time-evolving the representation and calculating expectation values can be computationally demanding.

Viewed more physically, the full many-body quantum state contains information about all possible correlators of observables (including multi-point correlators). This is important because even the attempt to predict only the evolution of single-particle observable expectation values in an interacting system will invariably require knowledge of higher-order correlators. This can be seen from the hierarchy of equations of motions that may be constructed for the set of all correlators and sometimes serves as a starting point for truncation schemes.

This is the challenge faced by a NN that has to learn to predict many-body dynamics based purely on observations. It has to construct a compressed representation that implicitly must involve information about the correlations that are building up inside the system due to interactions. Knowing only the single-particle observables and the current driving field would certainly not be enough to predict the subsequent evolution. We may, therefore, think of such a NN as discovering on its own, from scratch (without human guidance), a representation of a correlated many-body state that is hopefully more predictive than simple low-order truncation schemes. At the same time it should be also more general (e.g. in terms of dimensionality) than approaches based on ansatz wave functions. Moreover, this representation will be optimized for the set of observables that was selected for teaching the dynamics.


In this context, we mention another challenge. At first glance, it might seem obvious that once a NN has been taught to solve some representation of the effective equations of motion of the system, it can use this knowledge to predict the dynamics up to arbitrarily large times, exceeding the time spans seen during training. However, as we will explain below, we typically consider dynamics starting out from some simple initial quantum state. As time progresses, entanglement grows. This implies that the dynamics at later times, when longer-range correlations have built up, can be qualitatively different from that at early times, which makes extrapolation of the dynamics towards larger times a non-trivial task.

As we will see, two main ingredients have helped us to achieve our goal. First, as already alluded to, we teach the NN by having it observe the dynamics for time-dependent driving by many different realizations of a random process. This approach is simple but powerful, because it can immediately be applied to experimental setups. Second, in a systematic study of different NN architectures, we found that recurrent neural networks (RNNs), i.e., NNs with internal memory, excel at this challenge. These NNs are typically used for sequence processing and prediction, such as of audio signals \cite{graves2013speech} or in language translation \cite{sutskever2014sequence}. In the present context, they seem to become very efficient in time-evolving the autonomously discovered compressed representation of a quantum state.


\section{General remarks on the approach}

As indicated above, the main idea of our approach is to observe the system's dynamics under arbitrary driving, represented via time-dependent parameters $D(t)$ in the Hamiltonian. We train the NN such that it maps these driving trajectories (such as magnetic or electric fields evolving in time) to the observed evolution of the expectation values of a set of relevant observables. Specific classes of important driving scenarios include periodic driving or quenches, but the goal is to train a deep NN that is eventually able to provide accurate predictions for any sort of driving (such as what is encountered during quantum control protocols). As we will explain below, we achieve this key goal by training on driving trajectories that are samples of a stochastic process.

Furthermore, at this point, we want to stress that we train our NN to predict \textit{partial} observations, i.e., expectation values of only a selected subset of degrees of freedom, instead of a complete set of observables (i.e. the full quantum state). Because of this more modest, but practically very relevant goal, our approach is well-suited to be applied to experimental data, where a full quantum-state tomography during training would be infeasible. Moreover, the approach is compatible with a scenario where the internal dynamics of the system might even be unknown in the experiment, i.e. the Hamiltonian is not provided to us. Our NN is able to learn the dynamics while needing no prior information about the underlying dynamical equations. In contrast, standard methods based on, for example, exact Schr\"odinger evolution, time-dependent density matrix renormalization group \cite{schollwock2011density}, or more recent methods based on NN representations of quantum states \cite{carleo2017solving, gao2017efficient, schmitt2019quantum}, rely on precise knowledge of the general form of the underlying Hamiltonian ruling the dynamics of the device, whose parameters would need to be fitted to the experimental data. We note that there have been a few very promising recent first steps in the direction of training NNs on actual experimental measurement data, where the goal was to have the NN deduce the quantum state of a single qubit \cite{struck2020robust, flurin2020using}. To apply our approach in an experiment, the set of system observables would be measured at a given time and then averaged over many runs for the same driving trajectory in order to obtain an estimate of the expectation values. This procedure would be repeated for different time-points to obtain the full response for that driving trajectory.

Learning the dynamics from observations also immediately implies that the approach is not restricted to Hamiltonian dynamics, but rather allows the NN to correctly predict open and noisy quantum dynamics, i.e., dissipative dynamics that would have to be described either via Lindblad master equations or even via non-Markovian equations. Moreover,  while we will illustrate our approach further below on spatially homogeneous systems, inhomogeneous systems like impurity models or disordered systems fall within the purview of the method as well. By selecting suitable system observables, one can choose to focus on interesting spatial neighborhoods, e.g., the immediate vicinity of a quantum impurity.



As we shall illustrate for a specific example, our scheme provides substantial speedup as compared with direct numerical simulations of the Schr\"odinger equation. Exact Schr\"odinger evolution of a quantum many-body system involves an effort (memory and runtime) growing exponentially with system size.  In contrast, we will show that our approach provides a more favourable scaling in terms of system size (at most polynomial) for certain tasks that the full quantum state is not required. This is important for many applications, such as optimization (pulse engineering) and rapid exploration of the behaviour under different driving protocols \cite{PhysRevX.10.031002, PhysRevA.100.012103,RevModPhys.76.1037,RevModPhys.89.011004}.  Eventually, our approach could support efforts in autonomous design and calibration of complex quantum experiments \cite{melnikov2018active, moon2020machine}.

We briefly highlight another feature of our approach. As we will explain below, we rely on RNNs to predict the mapping from driving to observations. This has an important consequence: the NN can be used to produce predictions on time windows that are longer than what it has been trained on, and we will show that these are fairly accurate. This can find immediate applications for the extrapolation of experimental data in platforms with time constraints (e.g., existing quantum computers and quantum simulators with finite decoherence times). We will see that under appropriate conditions, our NN can even be used to extrapolate the dynamics to longer times for the infinite-system-size limit while it is trained on finite-size systems.

\section{Technical considerations \label{strategies}}

We employ supervised learning to train a NN on predicting the expectation values of the observables of interest, denoted by $S(t)$,  given the time-dependent driving trajectory $D(t)$. In this section, we comment on our specific choices for the key ingredients of the approach, that we have, after exhaustive explorations, found best suited to achieve high-quality predictions. These ingredients include the generation of training trajectories for $D(t)$ based on a random process, and the most promising combination of NN architecture and training strategy.

\textbf{Sampling Gaussian processes}.
We train the NN on arbitrary time-dependent trajectories $D(t)$ sampled from a sufficiently large class of functions, so that the NN can afterwards predict the resulting dynamics of expectation values $S(t)$ for different arbitrary time-dependent driving trajectories that it may have never seen during training. Investigating a few different classes of functions we noted that sampling these trajectories from a random Gaussian process works very well for our purpose. Our choice is motivated by well established properties of these distributions such as flexible shape in data fitting by controlling the mean and variance which made them the most fundamental distributions in statistics. In more detail, we use a mixture of Gaussian processes by randomly varying also their respective correlation times over a wide range. It turns out that in this way, the NN can learn all the essential features to finally predict the dynamics for different time-dependent functions. Sampling from a random Gaussian process for spatially random static potentials was demonstrated to be efficient in another context, for predicting band structures in topological materials \cite{peano2019rapid}.  There are various methods to generate Gaussian random functions \cite{liu2019advances}. In Appendix B. 1, we explain in detail the method that we use here (based on eigendecomposition of the correlation matrix). We also comment on the range of parameters (correlation time and amplitude of our Gaussian random processes) that we selected in our illustrative examples.

\textbf{Initial states}. It is important to realize that the fully trained NN will only be able to provide correct predictions for non-equilibrium quantum dynamics starting out in the class of initial states that was considered during training. 
We restrict ourselves to start with product states which are a common choice for most of real experimental implementations. However, our experiments confirm that our networks can also learn the dynamics for other classes of initial states given having as their input an appropriate description of the initial state.
 One can train the network on samples which are initially prepared in an identical entangled state or on samples for which the initial states are chosen from a specific class of  entangled states. For example, we observed that our network succeeds also in learning the dynamics  where the samples are  prepared initially in the ground state of the initial Hamiltonian. Note that the state preparation stage can also be included as part of the driving protocol during the prediction phase. This observation is inspired by the typical way that, e.g., an experimental quantum simulator would be run: At first, the system is prepared in a very simple state (e.g. a product state), and then an external pulse sequence or an adiabatic ramp of parameters is used to generate a more complex many-body state, which forms the starting point for all subsequent dynamics (e.g. a quench). We can follow the same idea here, albeit at the expense of using up part of the time interval during which the NN predictions are reasonably accurate. 

\begin{figure}[]
	\centering
	\includegraphics[width=1\linewidth]{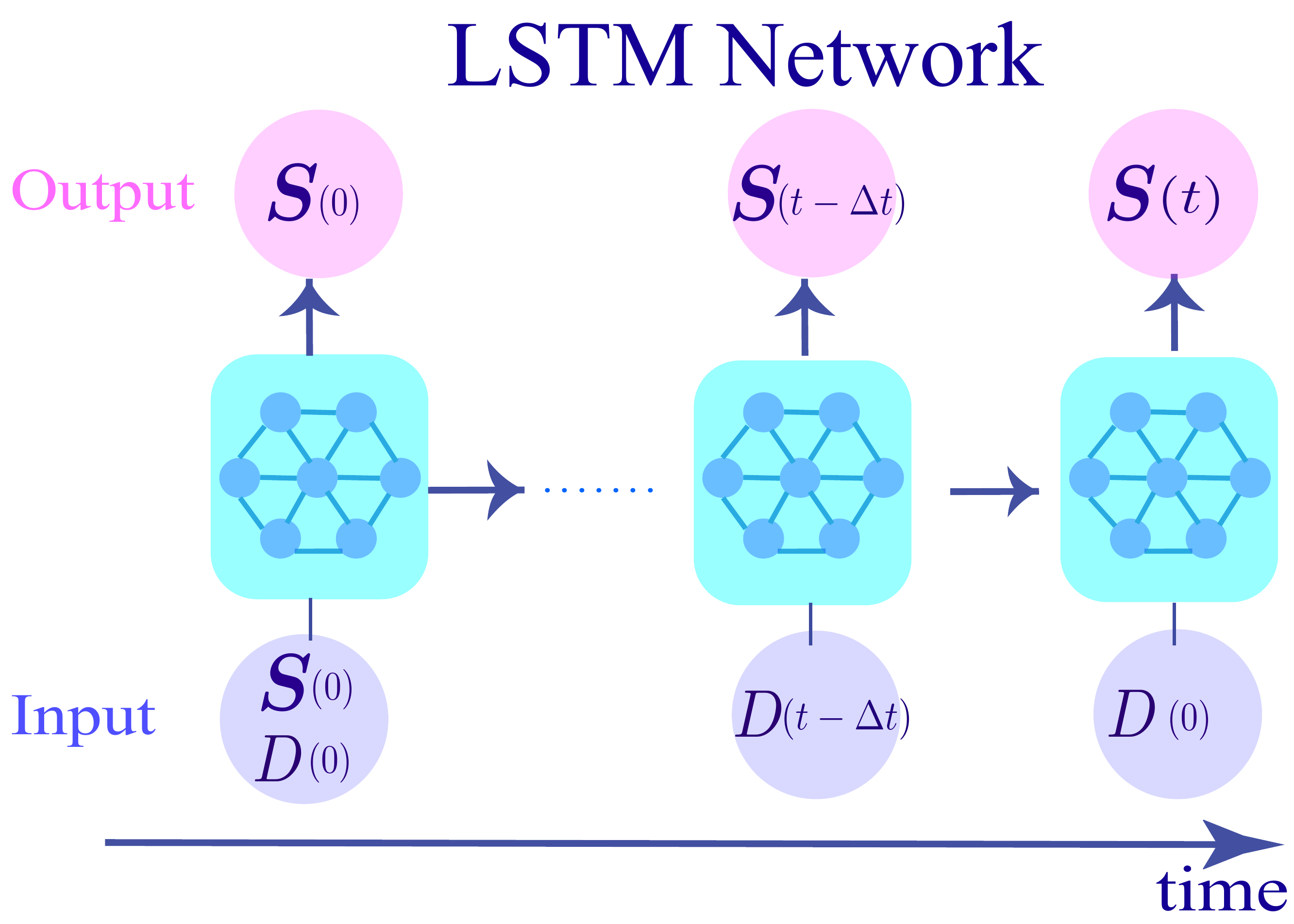}
	\caption {Schematic representation of our training strategy using a LSTM-NN. Horizontal arrows indicate the content of the internal neural memory being passed to the next time step.
	}
	\label{fig1_2}
\end{figure}

\textbf{General training strategy}. During training (Fig. \ref{fig1_2} (a)), we feed as input into the NN all the parameters that fully identify the Hamiltonian of the model, the entire driving trajectory $D(t)$, as well as a description of the initial state. Since we pick initial states $\rho(0)$ to be  a product state,   it is sufficient to provide a compressed unique description $f(\rho(0))$ of that state, which will not involve exponentially many parameters. In our illustrative examples, where we choose product states for the initial states, it is sufficient to provide the expectation values of a set of local observables (hence $f(\rho(0))=\boldsymbol{S}(0)$ in that case).   If one considers the case of an identical initial entangled state for all samples then it is not even required to feed to the network any information about the initial state.  The output of the NN is the full evolution of the expectation values of a subset of observables of interest ($\boldsymbol{S}(t)$). During training, in our case we provide the results of an exact simulation, but in future applications this data could come from an experiment.

\textbf{The neural network architecture}.
As for the architecture of the NN, we apply the most well-known type of RNN, ``long short-term memory" (LSTM \cite{hochreiter1735long}; Fig. \ref{fig1_2} (b)). As with any RNN, LSTM-NNs have a temporal structure and respect the fundamental principle of causality, which makes them well-suited to represent differential equations (equations of motion). Moreover, LSTMs specifically are able to capture both long-term and short-term dependencies. This characteristic is extremely useful as it gives the LSTM-NN the power to handle complex non-Markovian dynamics. Therefore, all of these together suggest this architecture to be very promising for our many-body dynamics prediction task (see Appendix  A for a brief recapitulation of the LSTM architecture).

Let us also comment here briefly on  the size of our LSTM-NN in terms of input and output size as well as number of hidden layers. The input size (at each time step) is  given by the number of driving  parameters  and  the number of parameters that identify our initial product state. The output size is set by the number of observables to be predicted. These observables may contain two-point or higher-order correlators. However, if we deal with a translationally invariant system and impose a cutoff for the range of all the correlators, the number of neurons in the output layer does not scale with system size. Even in the most general case (inhomogeneous system, arbitrary range of some finite-order correlators), the scaling is only polynomial.

The size of the hidden layers determines the number of trainable parameters in the NN. We will briefly comment in a later section on our empirical observations regarding the required size for the illustrative examples considered in this work (we find no need to scale up with system size).



\section{Physical Model\label{Model}}

To evaluate the performance of our scheme in predicting the dynamics of a driven quantum many-body system, we concentrate on two prototypical 1$D$ spin models (with periodic boundary conditions in each case): i) The transverse field Ising (TFI) model, where the spin ring is driven out of equilibrium with a spatially homogeneous time-dependent transverse field (Fig. \ref{fig1}(b)), and ii) the Heisenberg model, where we introduce a time-dependent coupling between neighbouring spins (Fig. \ref{fig1}(c)). Apart from their fundamental interest, these two models are of practical relevance in the context of non-equilibrium quantum many-body dynamics \cite{PhysRevB.98.024311, heyl2012nonequilibrium}. We discuss the Heisenberg model in Appendix I  and mostly focus on the TFI model here in the main text. 

The TFI Hamiltonian for a system of size $M$ reads
\begin{equation}
	\mathcal{H}_{\mathrm{TFI}}=B(t)\sum_{i=1}^{M}\sigma_{i}^{x}+J\sum_{i=1}^{M}\sigma_{i}^{z} \sigma_{i+1}^{z},
\end{equation}
where $\sigma_{i}^{\alpha}$ with $\alpha=x, y, z$ are Pauli operators acting on the site $j$. The spins form a one-dimensional ring, so that $\sigma_{M+1}^{\alpha} \equiv \sigma_{1}^{\alpha}$. $B(t)$ denotes the transverse magnetic field. The TFI model is a widely studied paradigmatic model with many experimental implementations. It displays non-trivial physics, with a quantum phase transition at $J=B$, and a rich phenomenology when driven out of equilibrium, in particular in the context of quenches \cite{chakrabarti2008quantum}. In the following, we set $J=1$ without loss of generality, since this only fixes the energy scale.

As the TFI model is a quantum-integrable model, one may wonder whether the success of the NN is tied to integrability. To clarify this we eventually also investigate non-integrable models, predicting the dynamics in the presence of an extra longitudinal field. In this case, the model cannot any more be mapped to free fermions and becomes non-integrable.

\textbf{Training}. We train the NN on Gaussian random fields $B(t)$ over a broad range of field amplitudes.  Specifically, for the numerical experiments analyzed here, we generated 50,000 random realizations of a Gaussian random process for $B(t)$. Later we analyze how the performance of the NN depends on the size of the training data set. We solve the Schr\"odinger equation for the TFI ring driven out of equilibrium by these random transverse fields, for the particular system size $M$ of interest. We calculate the evolution of a selected subset of observables. Note that in this work, for all the spin models that we investigate, we always choose as observables all the local spin operators $ \langle \sigma_{j}^{\alpha} \rangle$, as well as correlators of the type $\langle \sigma_{j}^{\alpha} \sigma_{j+\ell}^{\alpha'}\rangle$, with $\alpha, \alpha'=x, y, z$. This choice is for illustration only and does not imply one needs to train the NN over all these observables to get it to learn the dynamics properly. We verified that one can successfully train the NN on fewer observables of interest as well, see Appendix E for a detailed discussion on this.

We solve the Schr\"odinger equation for the case in which the spins are initially prepared in an arbitrary translationally-invariant uncorrelated state $\bigotimes_i(\sqrt{p}\vert 0 \rangle+\sqrt{1-p}\vert 1\rangle)$ with $p$ chosen at random from the interval $[0,1]$. Here, $|0\rangle$ and $|1\rangle$ denote, respectively, the $\pm 1$ eigenstates of $\sigma_z$. As we indicated above, starting from a product state does not restrict the applicability of our scheme, as one could start from more complicated states by including the state preparation as part of the driving protocol.

\begin{figure*}[th]
	\centering
	\includegraphics[width=1\linewidth]{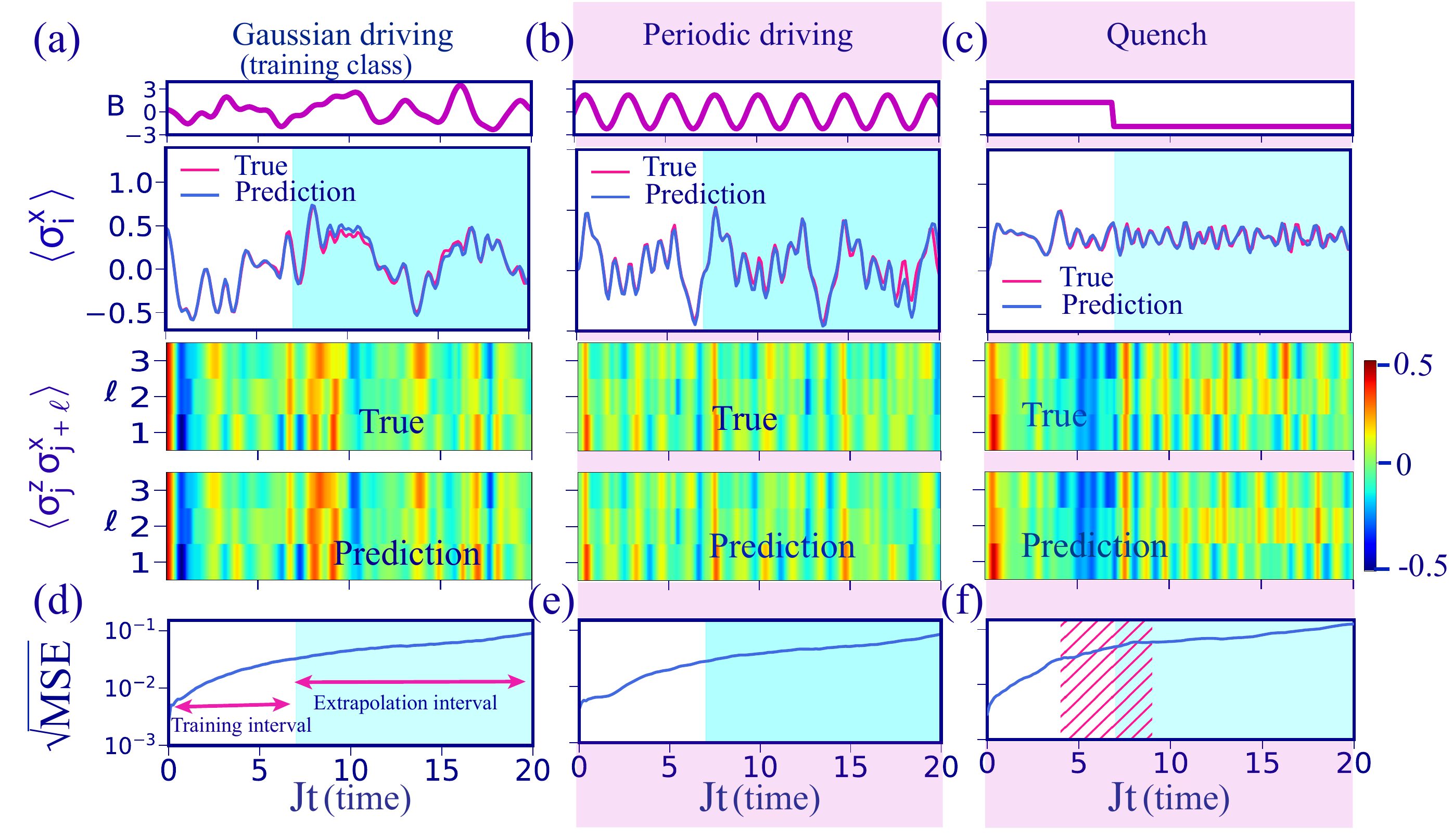}
	\caption{
		Demonstrating the high accuracy of the LSTM-NN in predicting the dynamics of a spin system under arbitrary driving. We train the NN on Gaussian random driving fields for a fixed time interval (here: $Jt\in[0,7]$) and then evaluate the prediction of the NN for different classes of driving fields, and also for a larger time interval ($Jt\in[0,20]$). In this example, we compare the true and predicted evolution of $\langle \sigma^x_i \rangle$ and $\langle \sigma^{z}_{i}\sigma^{x}_{i+\ell} \rangle$ for a transverse-field Ising spin ring of size $M=7$. Comparisons are displayed for different classes of driving fields: (a) a realization of a Gaussian process that has not been seen during training, (b) periodic driving, and (c) a quench. In the case of Gaussian driving, the spins are initialized in a translationally-invariant product state with random spin direction, while in (b) and (c) we chose to prepare the spins along the $z$-direction. The time window highlighted in light blue ($Jt\in[7,20]$) corresponds to a region beyond the one where the NN was trained on. The last row shows the evolution of the error (as given by $\sqrt{\textrm{MSE}}$, see main text) of the NN's prediction averaged over all the selected observables and over 1000 realizations of each type of driving field: Gaussian random fields in (d), periodic fields in (e), and (f) quenches between random static parameter values (quench times randomly chosen inside hatched region). For (d,e,f), the spins are initialized in a product state with random spin direction. Note that in the second and third column (highlighted in pink) the driving fields are not even in the class of functions that the NN has been trained on. 
	}
	\label{fig3}
\end{figure*}

\textbf{Evaluation}. In order to evaluate the performance of the fully trained NN and to show that it correctly predicts the dynamics for any type of time-dependent field, we test the NN on a new set of Gaussian random fields (with instances never seen during the training), as well as random periodic driving fields and different sorts of quenches. As for periodic drivings, we generate a set of functions of the form $B(t)=A \sin(\omega t) $ with random values for $A$ and $\omega$. As for quenches, we generate a set of step functions where the heights of steps and the time that the quench occurs is chosen randomly. See Appendix B. 2 for more details related to the range of parameters that we choose for generating our periodic and quench trajectories.


\section{Results}

In order to discuss the power of our approach in predicting the dynamics of a quantum many-body system, we inspect the prediction accuracy as a function of evolution time. 
We will also highlight two additional capabilities of our approach:  predicting the dynamics on a time window longer than those observed during training for (i)  the  system size that the NN was trained on, and  for (ii) the infinite-system-size limit, based only on training data for finite-size systems.


\subsection{Prediction accuracy}

For the sake of illustration, we consider a spin ring of size $M=7$ and train the NN on random Gaussian driving fields up to time $Jt=7$, where finite-size effects are already apparent. We then evaluate the performance of the NN for three classes of driving fields up to time $Jt=20$. Specifically, in the first column of Fig. \ref{fig3}, we evaluate the NN on a new set sampled from the the class of driving fields that the NN has been trained on, namely Gaussian random fields. In the next two columns (highlighted in pink), we consider some classes of driving fields that the NN has never seen explicitly, namely periodic driving fields and different sorts of quenches. In panels (a-c), we compare the predicted evolution for $\langle \sigma_j^{x}(t) \rangle$ and two-point correlators $\langle \sigma_j^{z}(t)\sigma_{j+\ell}^{x}(t) \rangle$ against their true evolution, for the driving fields shown at the top of each panel. The spins are initially prepared in random uncorrelated translationally-invariant states for the case of Gaussian driving fields, and we chose to initialize them along the $z$ direction for the other two types of driving fields. As can be appreciated, the LSTM-NN is able to predict the dynamics very well for the time interval that it has been trained on, and -- remarkably -- also to extrapolate to a longer time window (indicated by the light blue areas). We attribute both to the fact that the LSTM-NN has memory, so that it is able to build up by itself an implicit representation of the higher-order correlators using a non-Markovian strategy. We discuss this in more detail in Sec. \ref{H-O-C}. The results also confirm that training on Gaussian random driving fields is sufficient for the NN to learn all the essential features required for predicting the dynamics for arbitrary driving fields.

In order to show that the above-mentioned results hold beyond the particular driving trajectories that we displayed in Fig. \ref{fig3}(a-c), we present in Fig.~\ref{fig3}(d-f) the time evolution of the mean square error ${\textrm{MSE}}$. Throughout the paper, ${\textrm{MSE}}$ is defined as the quadratic deviation between the true dynamics and the predictions obtained from the NN, averaged over all samples and all the selected observables: ${\textrm{MSE}}=\langle|(\mathcal{O}_{j}(t)_{\rm NN}-\mathcal{O}_{j}(t)_{\rm true})|^2\rangle$, where $\mathcal{O}_j$ represents the expectation value of one of the observables of interest. In this case, we averaged over 1000 realizations of: Gaussian fields (d), periodic driving fields (e), and different sorts of quenches (f). In all cases, the spins are initialized in a randomly chosen, uncorrelated, translationally invariant state.

The light blue areas in Fig.~\ref{fig3} (and all figures of the paper) denote the time interval that the NN has not been trained on.   As one can see, also on this averaged level the predictions of the NN are valid even beyond the time window it has been trained on (remarkably, even for quenches that occur after the training time interval).

\begin{figure}[]
	\centering
	\includegraphics[width=1.00\linewidth]{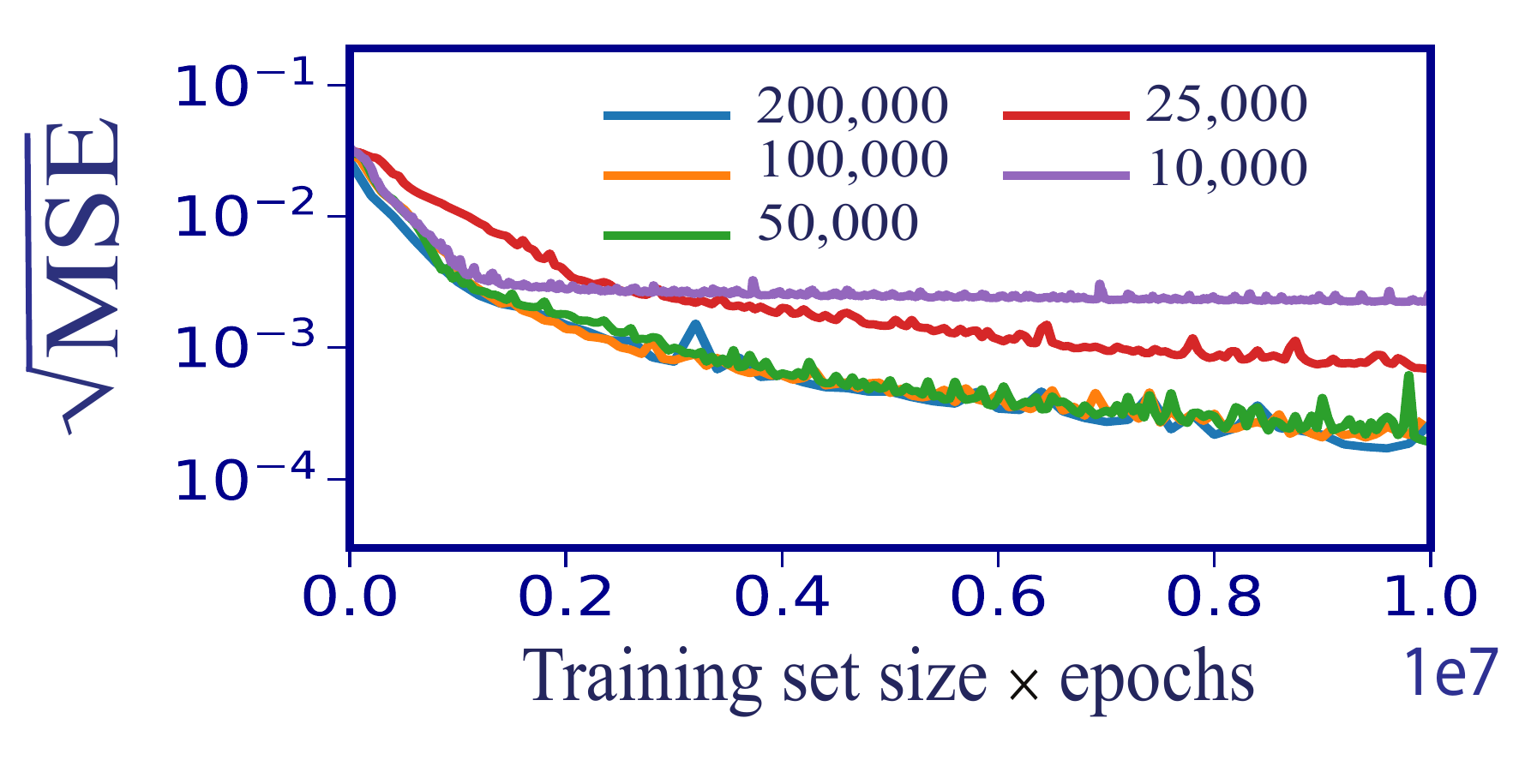}
	\caption{
		The error $\sqrt{\textrm{MSE}}$ versus number of instances that the NN has seen during training for different training set sizes (results for a $M=7$ spin ring,  training on Gaussian random driving fields).
	}
	\label{fig3_2}
\end{figure}

In Fig.~\ref{fig3_2}, we analyse how the prediction accuracy depends on the size of the training dataset. For this purpose, we plot for training sets of varying size the error $\sqrt{\textrm{MSE}}$ against the total number of training instances seen by the NN. As usual, training proceeds via epochs, each of which cycles once through the entire training data set. This plot verifies that 50,000 samples are already sufficient to achieve high accuracy, but smaller training set sizes, down to 10,000 samples can already achieve reasonable learning success. This may be important for experimental implementation, where generating training data can be time-consuming, depending on the runtime for an individual shot of the experiment. We refer the reader to  Appendix D for a more detailed discussion of the resources required to train the NN in terms of data set size and NN size.

Let us to point out that, in general, for stronger transverse magnetic fields, it becomes more challenging for the NN to approximate the dynamics over a longer time window. We attribute this to the fact that for very large transverse fields, the expectation values of spin operators oscillate quite rapidly. 
Still, as can be seen from the plots above, the prediction of the NN for the range of driving amplitudes that we have considered here (see Appendix B, for more details) most of the time remains valid up to $Jt =20$, substantially longer than the training time interval.

\subsection{Extrapolating the dynamics to the infinite-system-size limit}

We now turn to the question whether one could train a NN on a finite-size system and yet have it extrapolate the dynamics to the infinite-system-size limit. We have found that this goal can be achieved by training the NN for a sufficiently large system size up to a point in time where the finite-size effects have not yet appeared. As we will show in this section, once the NN (trained in this manner) is asked to extrapolate the dynamics to longer times, the predictions fit the infinite-system-size-limit dynamics.

In our example, we train the NN on system size $M=15$, which is the largest system size for which we are still able to generate training samples in a reasonable time by solving the many-body Schr\"odinger equation. For a given system size, the time window for which the observed dynamics is effectively indistinguishable from the infinite-system-size limit depends on the chosen driving field and also on which observable we are monitoring. Moreover, we have discovered an even stronger dependence on the initial state of the spins. For example, for system size $M=15$, given initial spin states closely aligned with the magnetic field (paramagnetic initial condition), finite-size effects appear quite early, say for $Jt\in[3,15]$. However, if we initially prepare the spins in other directions, finite-size effects appear for $Jt\in[7,25]$ for most observables. Therefore, in the following, we present results for initial paramagnetic states, with the NN trained up to $Jt=3$, while for any other type of initial states training occurs up to $Jt=7$.
\begin{figure*}[!ht]
	\centering
	\includegraphics[width=1.00\linewidth]{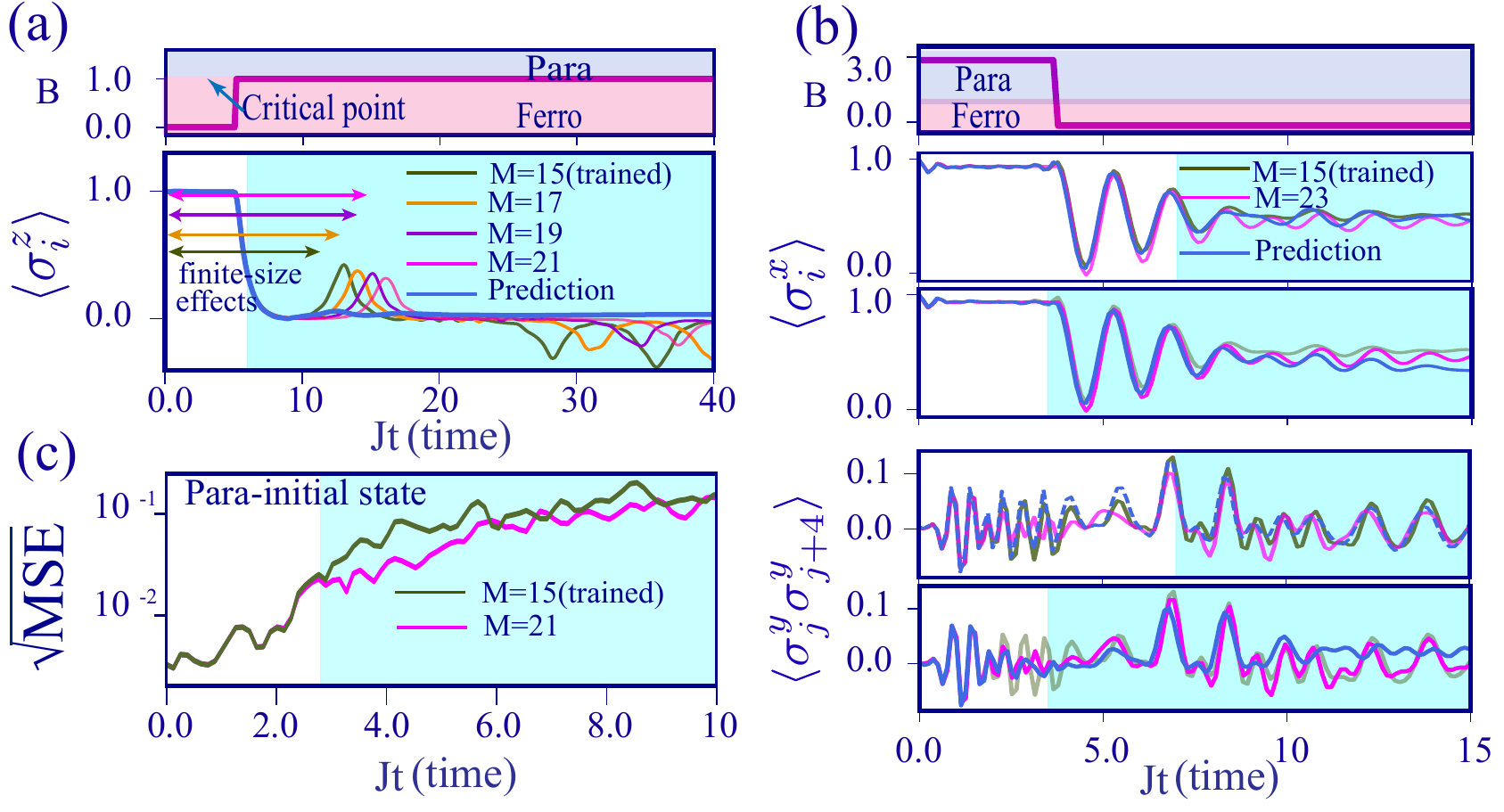}
	\caption{
		Power of the LSTM-NN in extrapolating the dynamics to the infinite-system-size limit. The NN is trained for a TFI spin ring of size $M=15$, on Gaussian random driving fields, in the time interval marked in white. We compare the NN prediction for longer times with the true dynamics of larger system sizes, as indicated in the figures.
		(a) Quench from a ferromagnetic phase to the critical point (spins are initially prepared in a ferromagnetic state).
		(b) Quench from a paramagnetic phase to a ferromagnetic phase (spins are initially prepared in a paramagnetic state). For this case, and for the purpose of comparison, we present results for two different training time windows, $Jt\in[0,7]$ and $Jt\in[0,3]$, where finite-size effects are already visible or not, respectively (see main text).
		(c) The time-evolution of the prediction error, comparing the NN predictions against the true dynamics for $M=15$ (green curve) and $M=21$ (pink curve), averaged over 30 randomly sampled quenches. Spins are initially aligned in a paramagnetic initial state. 
	}
	\label{fig4}
\end{figure*}

In Fig.~\ref{fig4}, we illustrate the power of our LSTM-NN in extrapolating the dynamics to the infinite-system-size limit, under different sorts of quenches. To verify the power of our NN in extrapolating to this limit, we compared the NN prediction against the largest system size ($M=21$) that we have still been able to solve using the exact Schr\"odinger evolution (for the particular TFI model, we could also have resorted to the known exact solution, but in general, for abritrary models, that would not be possible). 
We investigated two scenarios:

(i) Quench from a ``ferromagnetic'' phase to the ``critical point'': the spins are initially prepared in a ferromagnetic state $\bigotimes_i|\uparrow\rangle$, and we quench to the critical point ($B$ is suddenly changed from $0$ to $B=J$). In Fig \ref{fig4}(a), we show the power of the NN in predicting and extrapolating $\langle \sigma^z _i\rangle$ under this quench. As in the previous figure, the region highlighted in light blue corresponds to the times beyond those where the NN has been trained on. We show the true dynamics for different system sizes in order to appreciate how finite-size effects (here in the form of revivals) appear later for larger sizes. Since we train up to a time where finite-size effects have not yet appeared, the NN is able to extrapolate the dynamics to the infinite-system-size limit, as is apparent from the plot (it shows no revivals).


(ii) Quench from a ``paramagnetic'' phase to a ``ferromagnetic'' phase: the spins are initially prepared in the paramagnetic state $\bigotimes_i |\rightarrow\rangle$, where $|\rightarrow\rangle$ is the $+1$ eigenstate of $\sigma_x$, and we quench into a ferromagnetic phase ($B$ is suddenly changed from $2.8J$ to $-0.1J$). We present the performance of the NN for the cases where it was trained up to $Jt=7$ and $Jt=3$, for which the finite-size effects are already present or not, respectively. As can be seen, when training is performed on the longer time interval, the NN predictions  match the evolution for the system size that it has been trained on ($M=15$ in this case), including the corresponding finite-size effects. Remarkably, for a \textit{shorter} training time interval where finite-size effects are not yet visible, the NN predictions are closer to the \textit{larger} system size $M=23$, approaching the infinite-size limit.

This behaviour can also be observed when averaging over many different trajectories. In general, we observe that when we initially prepare spins in the paramagnetic initial state (i.e. aligned along the $x$ direction), it is more difficult for the NN to extrapolate the dynamics to the infinite-system-size limit. More specifically, in that case the extrapolation to the infinite-system-size limit is valid for a shorter time window. We attribute this to the fact that finite-size effects appear quite early in this case. We checked out different sorts of quenches with the spins in a (i) ferromagnetic (aligned along $z$) and in a (ii) paramagnetic initial state. For each class, we evaluated the NN on 30 realizations. We observed that the NN is able to extrapolate  for a longer interval and with a higher precision for the ferromagnetic initial state. In Fig. \ref{fig4}(c), we show $(\sqrt{\textrm{MSE}})$ for the (more challenging) paramagnetic initial state. The green(pink) curve shows the quadratic deviation of the NN prediction from the exact evolution for $M=15$ ($M=21$).  As one can see, the NN predictions are closer to the larger system size $M=21$ than the system size that it has been trained on ($M=15$) verifying the power of the NN in extrapolating to infinite-system-size limit. After a while, both curves approach each other, which is due to the fact that finite-size effects for system size $M=21$ start to appear as well.

\subsection{Implicit construction of higher-order correlators by the neural network\label{H-O-C}}

Even though we are only interested in predicting the dynamics of single-spin observables and two-point correlators, it is a well-known fact that the dynamics of those depends in turn on higher-order correlators. For example, in the particular model (TFI spin ring) discussed above, these equations of motion are:

\begin{equation}
	\dfrac{d}{dt}  \langle \sigma^n _{i}\rangle=-2 \sum_{k}\left[B(t) \epsilon_{xnk}  \langle \sigma^k_{i} \rangle+2 \epsilon_{znk}  \langle \sigma^z_{i} \sigma^k_{i+1} \rangle\right],
\end{equation}

\begin{align}
	\dfrac{d}{dt}  \langle \sigma^m _{j} \sigma^n _{j+\ell}\rangle=&
	\\
	&\hspace{-1cm}-2\sum_{k}\big[B(t)( \epsilon_{xnk}  \langle \sigma^m_{j} \sigma^k_{j+\ell} \rangle + \epsilon_{xmk}  \langle \sigma^k_{j} \sigma^n_{j+\ell} \rangle)\nonumber
	\\
	&\hspace{-0.8cm}+\epsilon_{znk}(\langle \sigma^m_{j} \sigma^k_{j+\ell}\sigma^z_{j+\ell+1} \rangle+\langle \sigma^m_{j} \sigma^z_{j+\ell-1}\sigma^k_{j+\ell} \rangle)\nonumber
	\\
	&\hspace{-0.8cm}+\epsilon_{zmk}(\langle \sigma^z_{j} \sigma^k_{j+1}\sigma^n_{j+\ell+1} \rangle+\langle \sigma^k_{j} \sigma^z_{j+1}\sigma^n_{j+\ell}\rangle)\big],\nonumber
\end{align}
where $n, m, k \in {x,y,z}$ and $\epsilon_{ijk}$ denotes the Levi-Civita symbol. 

\begin{figure}[t]
	\centering
	\includegraphics[width=1\linewidth]{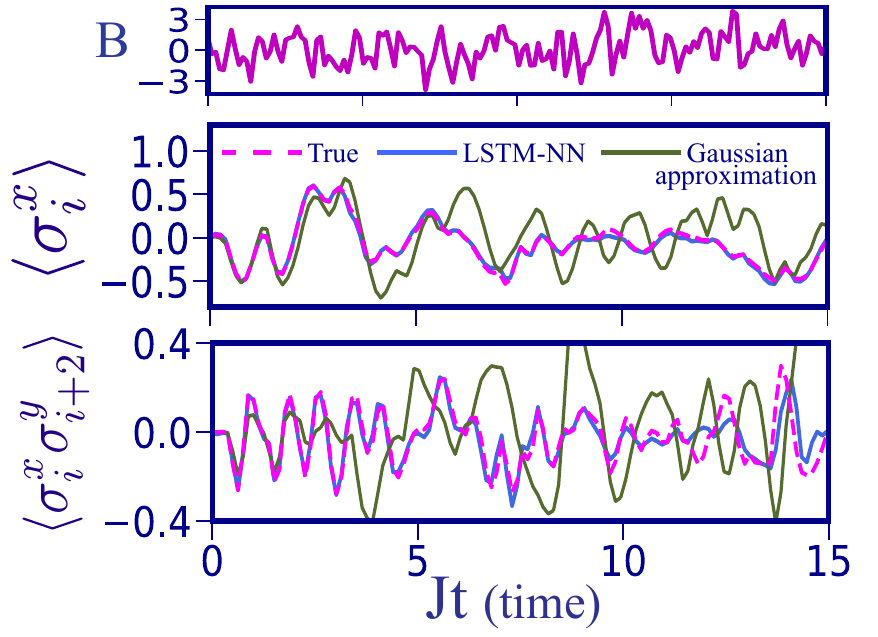}
	\caption{
		Illustrating the power of the NN in finding internal representations of higher-order correlations. We compare the NN predictions with the solutions of the equations of motion for correlators, truncated at the Gaussian level. The NN prediction is much closer to the true dynamics, especially at longer times. (A spin ring of size $M=7$ is driven by the transverse field shown at the top of the panel; spins are initially prepared in the $z$ direction)
	}
	\label{fig6}
\end{figure}

The fact that the NN is able to approximate the dynamics with high accuracy for long times even though it has not been trained on predicting higher-order correlators implies that it is able to build up some internal representation of the latter by itself. In order to verify this, we compare the NN predictions against the well-known 'Gaussian' approximation that is obtained by truncating the equations of motion at the level of two-point correlators. We apply the Gaussian moment theorem \cite{mandel1995optical} (for arbitrary Gaussian operators $A$, $B$, and $C$) to the third-order moments as follows
\begin{align}
	\langle ABC \rangle=\langle AB \rangle \langle C \rangle+ \langle BC \rangle \langle A \rangle+ \langle AC \rangle \langle B \rangle- 2 \langle A \rangle \langle B \rangle \langle C \rangle.
\end{align}
 and obtain a closed system of equations for the first and second-order moments.  In Fig.~\ref{fig6}, we compare the evolution of $ \langle \sigma_{i}^{x} \rangle$ and $ \langle \sigma_{i}^{x} \sigma_{i+2}^{y}\rangle$ obtained from these truncated equations, from the NN, and from the exact dynamics. As can be seen, the dynamics extracted from the Gaussian approximation is valid only during a short time, while the prediction of the NN holds during much longer times. This verifies that the NN is successful in finding an implicit representation for higher-order correlators using a non-Markovian strategy (memory). To illustrate that this is generally true, we provide a much more detailed comparison in  Appendix F).

\begin{figure}[t]
	\centering
	\includegraphics[width=1\linewidth]{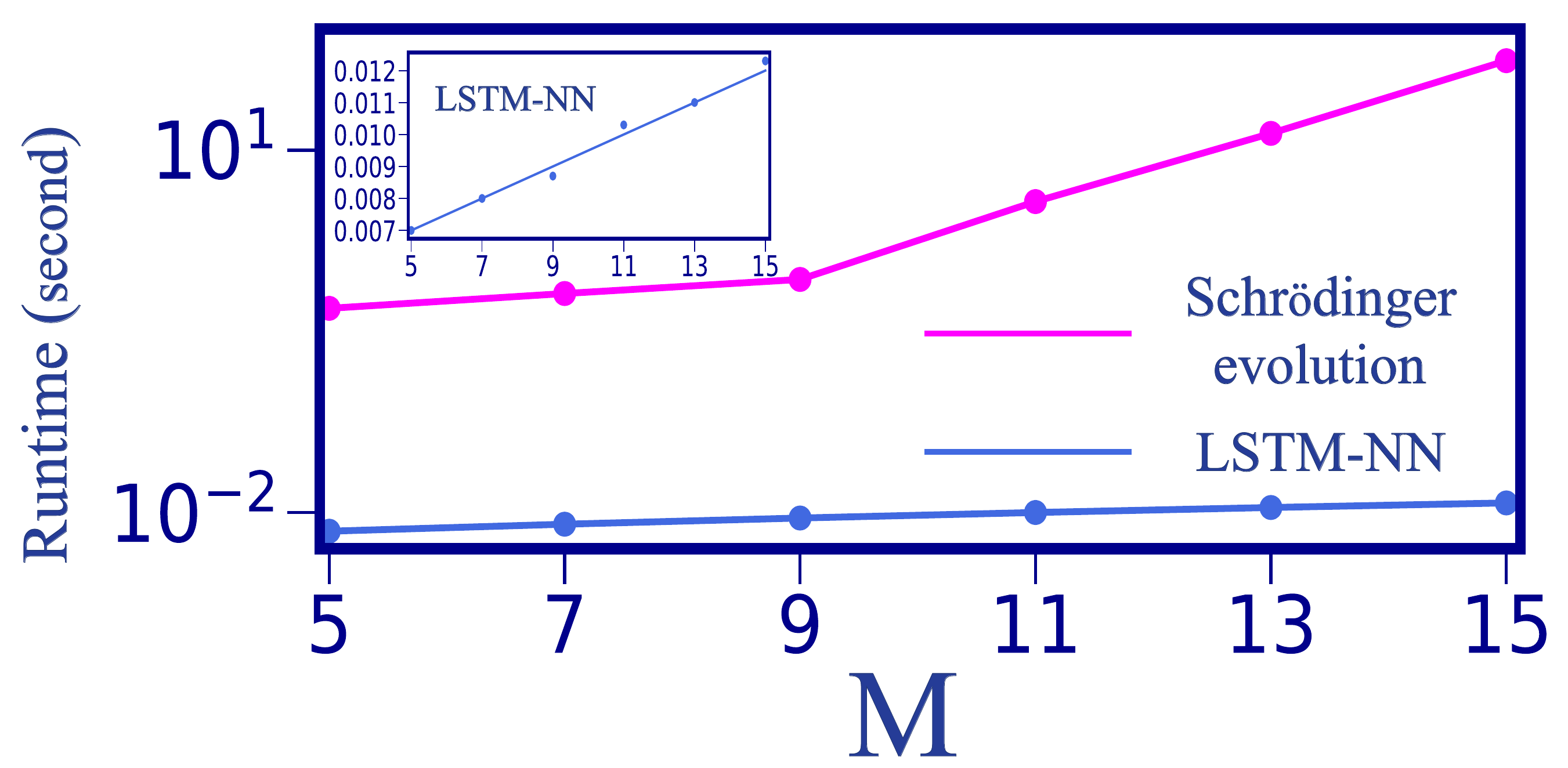}
	\caption{
		Runtime versus system size, comparing the time needed for evaluating the NN against direct Schr\"odinger evolution (evaluated using the qutip implementation). The inset shows the runtime  versus system size for the NN in normal scale. The shown results are for TFI model driven by  Gaussian random fields up to $Jt=20$.
	}
	\label{fig7}
\end{figure}

\subsection{Speedup}

We recall that, as mentioned before, our approach offers advantages that go much beyond any possible speedups vs. direct numerical simulations, such as, in particular, the ability to learn the dynamics of an unknown experimental system. However, let us still comment on the speedup that can be achieved. 

The speedup achieved by our trained NN depends on the algorithm which it replaces, as well as the particular implementation. As we do not put any constraints on the dimensionality of the physical system, or any other aspects, the natural point of comparison is the direct simulation of the Schr\"odinger equation (rather than powerful approximate methods like matrix-product states that are more restricted in their use). Of course in this work we just explored models in one dimension. However, when moving to higher space dimensions typically systems thermalize more easily via interactions, which means the evolution of observables can become even simpler (despite the large growth of entanglement, presenting a problem for wave-function-based methods). When that is the case, the NN should be able to learn even more easily.

In setting up the comparison between the NN and a direct numerical integration of the Schr\"odinger equation, we have to be careful. It would be unfair to impose the default (very stringent) numerical accuracy requirements on the numerical integrator, since the NN only gives approximate predictions. For that reason, we relaxed the accuracy setting until the numerical integrator had errors on the order of the typical deviations observed in the NN. In addition, we allowed adaptive stepsize for increased efficiency (See Appendix G for more details). Besides, we selected the sampling step, for which the integrator returns the expectation values, as large as possible. The NN  is asked to predict the expectation values on a grid that is twice as fine.

Any direct Schr\"odinger evolution of a quantum many-body system involves an effort (memory and time) growing exponentially with system size, which is avoided by the NN approach as one can see in Fig. \ref{fig7}. In this figure, we compare how the runtime scales versus system size $M$, comparing a single evaluation of the NN with the direct Schr\"odinger evolution up to time $Jt=20$. Both algorithms were run on the same CPU, and the numbers given are averaged over 1000 instances.  As can be seen, the runtime  for the NN scales linearly versus system size (see inset). Note that here we trained the NN on predicting all spatial two-point correlators of the spin operators, meaning that the size of the output layer of the NN increases linearly with system size. We do not see any reason that prevents such a favourable scaling for larger system sizes. Such polynomial scaling should even hold for inhomogeneous models.

Despite our efforts to account for aspects of accuracy, the speedup comparisons made above may still seem unfair at first glance.  First, we have not included the effort for training here. Second, the NN is only able to predict the dynamics for the subset of observables it has been trained on; in contrast, the Schr\"odinger evolution provides access to the full quantum state. Concerning the first point, we argue that there are many practically relevant applications which demand exploration of a much larger number of samples than those required to train the NN, and in this case the speedup by the NN eventually outweights the training effort. Examples include rapid exploration of many different parameter values or driving protocols, and pulse engineering, where the goal is to find the driving trajectories that lead to a desired behaviour of the system.  Regarding the second point, we emphasize that there is an important class of applications where we are interested in the dynamics of a subset of observables and not in the full quantum state. In addition, in any typical experiment, only such a subset of observables is accessible in the first place. All of these aspects taken together make the comparison introduced here relevant for practical applications.

To provide some illustrative numbers (with all the caveats regarding dependence on computer hardware and algorithm), for a single instance at system size $M=15$, the LSTM-NN is able to predict the dynamics up to time $Jt=20$ in 15 ms, while solving the Schr\"odinger equation directly for the same instance takes about 16s. Thereby, not taking into account the training effort, this particular example shows a remarkable speedup of a factor $1000$.



Although we have not performed a systematic study of how the hidden layers should scale with system size to reach a certain accuracy, we can comment on what we infer from our observations on the models that we explored. For the translationally invariant models (with a cutoff for the range of correlators), the number of hidden layers does not scale up with system size.  A more complicated scenario, that we do not explore in this work, is for spatially inhomogeneous models where the number of output neurons can naturally scale linearly with system size. For such cases, one needs to apply a convolutional neural network (CNN) to deal with the spatial dimension, combined with the LSTM-NN that tracks the time evolution. Then even though the number of output neurons  scales (linearly) with system size, the nature of a CNN implies that the number of trainable parameters (weights) is still independent of system size. The only exception would be at or near a critical point, where very long-range correlations develop, which may make it necessary to grow the CNN weight kernels (or number of layers) with the system size, but not stronger than linear.

We also point out that NN training can exploit parallelism with respect to static parameters in the Hamiltonian. This means if a NN has been trained on a full range of these values (while supplying the parameter values as additional input), the result will be a NN that makes better predictions than what would have been obtained when using the same overall amount of training samples to train several separate NNs, each for a specific parameter value (or a specific small parameter interval). The reason is that the ``meta-network" (that has been trained on all different parameter values simultaneously) effectively re-uses what it has learned for one parameter value to improve its predictions at other values.

\subsection{Predicting the dynamics of quantum non-integrable models \label{integrability}}

\begin{figure*}[]
	\centering
	\includegraphics[width=1\linewidth]{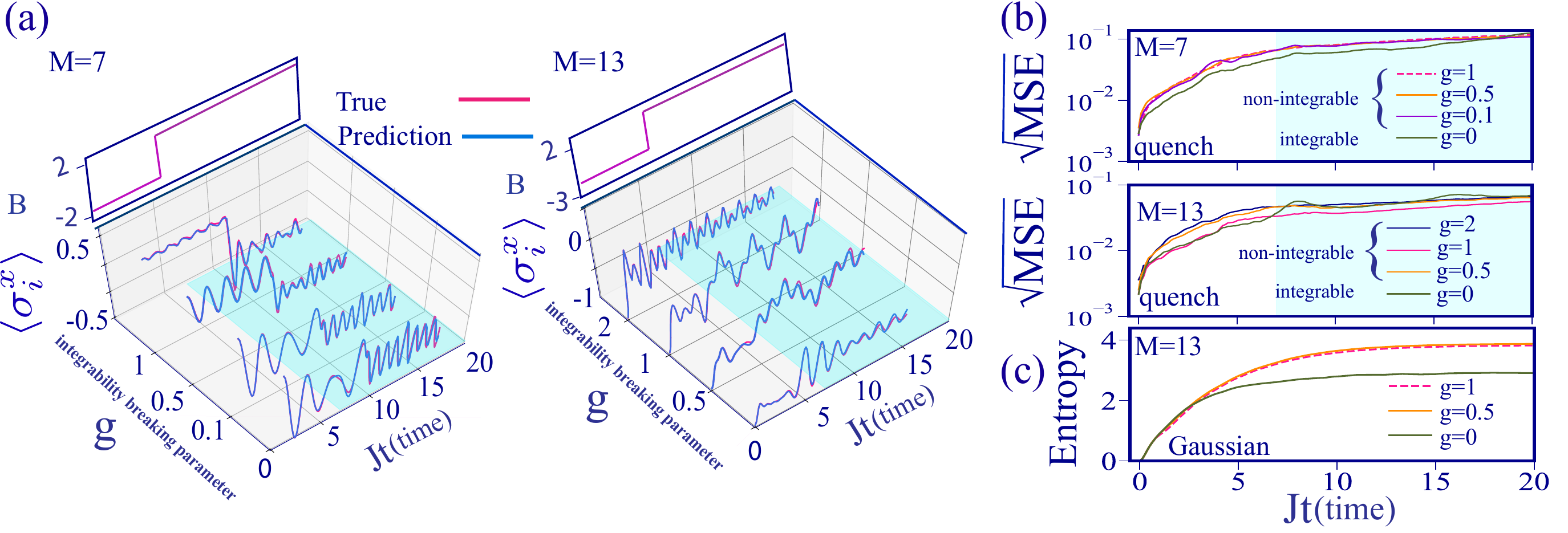}
	\caption{
		Comparing the power of the LSTM-NN in predicting the dynamics of quantum-integrable models versus non-integrable models.
		(a) Comparing the performance of the LSTM-NN in predicting $\sigma_{i}^{x}$ against its true evolution for different values of the integrability-breaking parameter $g$ under the quenches shown here. 
		(b) Prediction error versus time, for different values of $g$, where the spin-ring is subject to different sorts of quenches. For each value of $g$, the NN is evaluated on 1000 realizations for system sizes $M=7$ and $M=13$. The light blue regions show the interval that the NN has not been trained on. (c) Evolution of the entanglement, as measured by the von Neumann entropy of a subsystem (averaged over 1000 realizations of random driving fields).
	}
	\label{fig8}
\end{figure*}

The TFI model with an arbitrary time-dependent transverse driving field is a quantum-integrable model \textit{at every instant of time}. This means that at each time, it can be readily diagonalized by a linear transformation of the fermionic operators that result from applying the Jordan-Wigner transformation \cite{mbeng2020quantum} to map this spin model to a quadratic fermionic model. Thus, one might ask whether the success of the NN predictions is due to the integrability. In order to rule out this potential explanation, we discuss the power of our NN for a more general, non-integrable Hamiltonian as well. We add a longitudinal field to the TFI Hamiltonian,
\begin{equation}
	\mathcal{H}=B(t)\sum \sigma_{i}^{x}+g \sum \sigma_{i}^{z}+J \sum \sigma_{i}^{z} \sigma_{i+1}^{z},\label{noninteg}
\end{equation}
which is known to break integrability.

We are interested in investigating how the power of the NN in predicting the dynamics is related to the strength of the integrability-breaking parameter $g$. Thus, we train the NN on the above Hamiltonian with  Gaussian random transverse driving fields for a fixed value of $g$ and evaluate its performance for the same $g$ on a different class of driving fields. We eventually repeat this comparison for a set of different $g$.


In Fig.~\ref{fig8} (a), we compare the predicted and true evolution of $\langle \sigma_{i}^{x} \rangle$ under specific shown quench trajectories, for different values of $ g $.  Apparently, the NN is able to predict the dynamics successfully both for the time-window (white regions) that it has been trained on and beyond that (light blue regions).

For a more quantitative analysis, we show in Fig.~\ref{fig8} (b) the prediction error versus time, for different sorts of quenches. For each coupling $g$, we average over 1000 realizations and over all our observables. We note that, in general, it is not fair to compare the error for models with different values of $g$, as they have completely different dynamics. But at least Fig.~\ref{fig8} (b) demonstrates that the NN also succeeds in predicting the dynamics of the investigated non-integrable models, with various different integrability breaking parameters $g$. Nevertheless, as can be seen in this figure, there is a tendency for the NN predictions to be less accurate for the non-integrable models. Evaluating the NN on more classes of driving fields, we observed that less accurate predictions are more evident for smaller system sizes  at long times. In such cases the NN has a tendency to predict trajectories close to the infinite-system-size limit, rather than the given system size that it has been trained on. At this moment, it remains unclear whether this is a consequence of the model being non-integrable. It is also unclear whether this would hold across a wider range of physical models, as we observed the same behaviour for the Heisenberg model with time-dependent exchange couplings, which is known to be  quantum integrable at each instant. In  Appendix H, we present some additional analysis, for Gaussian random driving fields. 


Moreover, in Fig.~\ref{fig8} (c), we show how entanglement grows in terms of time for the model (\ref{noninteg}). Specifically, we calculate the von Neumann entropy of half the spin system. As it is evident, the entanglement already grows significantly in the time window for which the NN successfully predicts the dynamics (both within the training interval and beyond that). This again verifies the power of the NN in constructing an implicit representation of complicated entangled quantum many-body states,  in terms of higher-order correlators, as already discussed in  Sec.~\ref{H-O-C}.

\section{Other training strategy and other neural network architectures\label{archs}}

All our analysis so far has relied on the LSMT-NN that has to predict the observables when receiving the driving trajectory as input. To put these results into perspective, we have analyzed a variety of other approaches, including other NN architectures and another training strategy, which we will now discuss.

\begin{figure}[]
	\centering
	\includegraphics[width=0.75\linewidth]{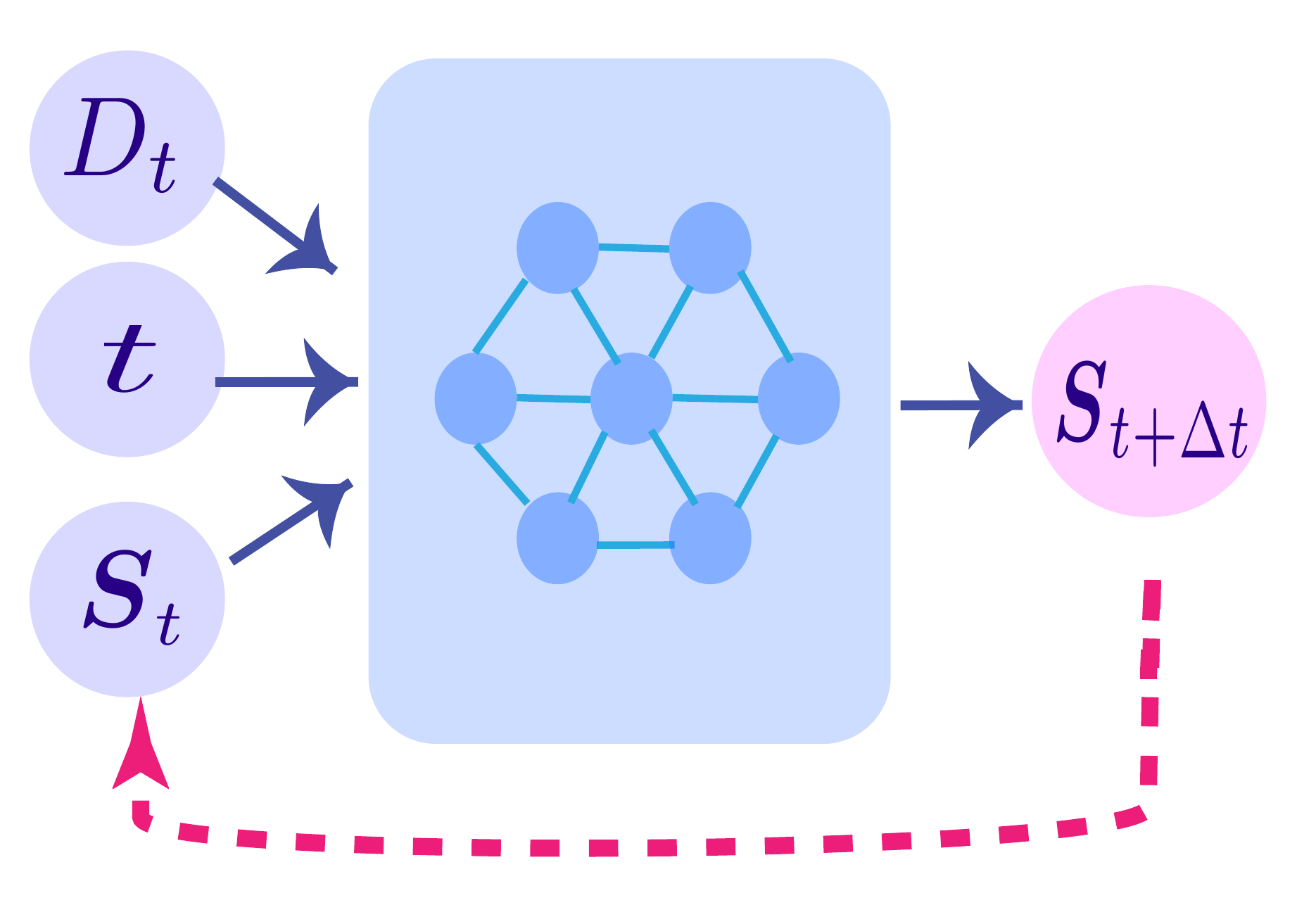}
	\caption{
		Schematic representation of the alternative training strategy (``step-wise training'') introduced in the text. The output of the NN (i.e. the predicted evolved expectation values) is fed back as input, generating the dynamics one step at a time.
	}
	\label{fig9}
\end{figure}

As an alternative to the training strategy introduced in Sec.~\ref{strategies}, which we will further refer to as ``full-evolution training'', we now discuss an alternative training strategy that we can apply for our many-body dynamics prediction task. We call  this strategy ``step-wise training''. In this strategy, we ask the NN to learn a set of Markovian evolution equations (Fig. \ref{fig9}). We start by feeding into the NN the initial value of the observables of interest, together with the initial value of time and the value of the driving field. We train the NN to predict the value of the observables for one time step ahead. Moving on to the next time step, we feed into the NN the previously predicted observables, together with the current value of time and the driving field. The NN then predicts the observables for the next time step ahead, and we keep going until we reach the last time step. We stress that this strategy explicitly prevents the NN from directly building up any internal representation of higher-order correlators as long as it is not recurrent, so it represents a useful benchmark. Of course, the NN could still try to approximately infer some higher-order correlators from the observables which it is supplied with as input at any given time, provided that the dynamics does not explore completely arbitrary states.

In addition to applying both of these training strategies, we also explore other types of architectures for the NN. For example, inspired by the time-translational symmetry of the dynamics together with the fact that CNNs can be made to respect causality and can also effectively represent memory effects, we discuss the power of a temporal 1D-CNN in predicting the dynamics. Moreover, we explore the power of a fully-connected neural network (FCNN) on this task. 

\begin{figure}[h]
	\centering
	\includegraphics[width=1\linewidth]{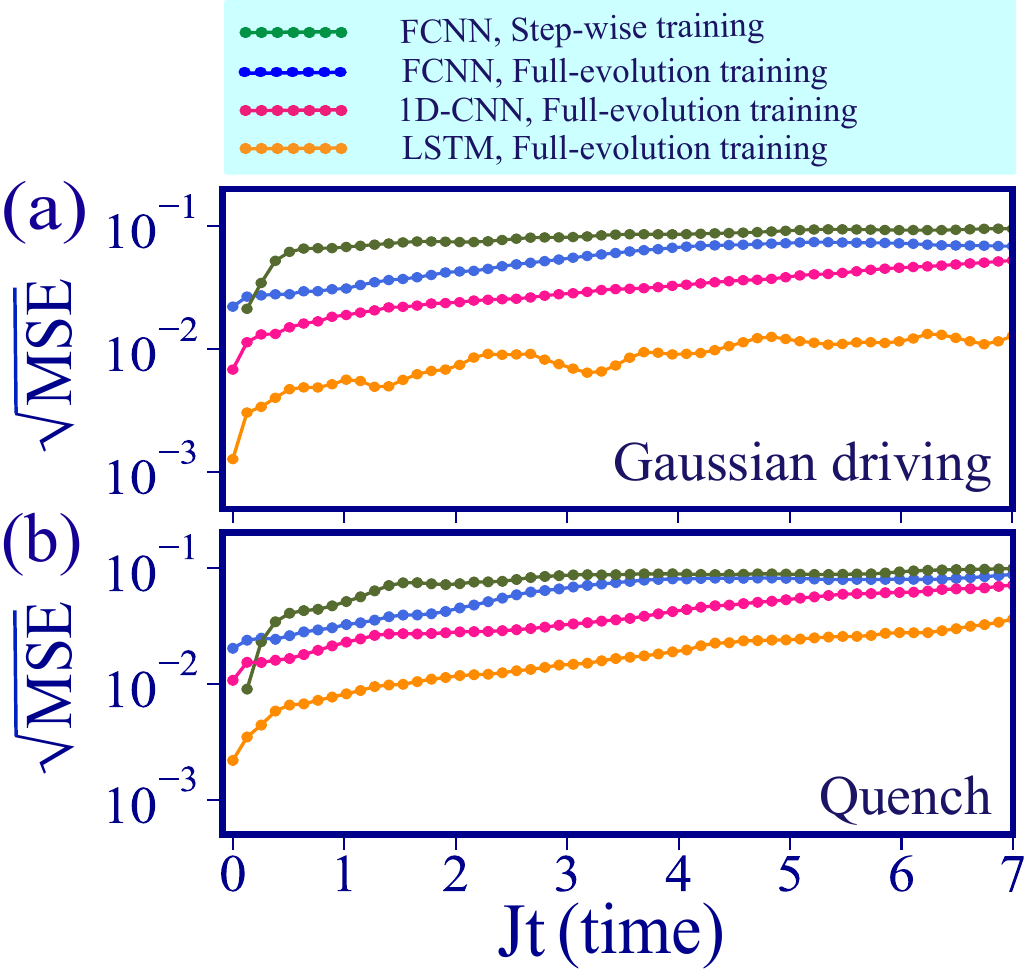}
	\caption{
		Comparing the power of different training strategies and NN architectures in predicting the dynamics of an arbitrarily driven spin system. We show the prediction error versus time (TFI spin ring of size $M=15$), for driving by (a) 1000  Gaussian random driving fields and (b) 1000 different sorts of quenches. For all cases, the spins are initially prepared in a translationally invariant product state with random spin direction. The combination of LSTM-NN and full-evolution training is clearly superior to all other approaches.
	}
	\label{fig10}
\end{figure}

Next, we compare the performance of all of these training strategies and architectures by concentrating on a TFI model with system size $M=15$. We train the NN up to $Jt=7$ on  Gaussian random driving fields, and evaluate all architectures on the same time interval on a new set of  Gaussian random driving fields (Fig. \ref{fig10} (a)) as well as different sorts of quenches (Fig. \ref{fig10} (b)). For each class, we averaged over 1000 realizations, and over all the selected observables.
For all samples, the spins are initially prepared in a random translationally-invariant product state.

Among different approaches, it is clear that the previously analyzed combination of the ``full-evolution training strategy" and the LSTM-NN is superior. Notably, this scheme also outperforms the 1D-CNN, that can in principle apply a non-Markovian strategy to capture the dependencies on the past at each time. We attribute this to the fact that the LSTM-NN is able to decide in each step in a systematic way which long-term memories to keep. In more detail, it is able to capture both short- and long-term dependencies, while this is not the case for the CNN, where the choice of temporal kernel size also restricts the memory time span (see Appendix C. 2 for details related to the layout of 1D-CNN architecture, the optimal kernel size, and also  error  versus time for different kernel sizes).

The green and dark blue curves in Fig.~\ref{fig10} compare the power of a FCNN applying both training strategies. For the step-wise training, the error grows faster at the beginning. This may happen for two reasons. First, in each step the NN has access to the information from just one step ago. Therefore, it does not have a chance for building an implicit representation for the higher-order correlators that develop with time. Second, in each step we feed to the NN predictions from one step ago, which leads the NN to accumulate error at each step. For all of these reasons, the ``full-evolution training'' strategy looks more promising and indeed performs better.

It is impressive that an FCNN applying the full-evolution training strategy still has a reasonable performance up to some point, even though it has no built-in notion of causality and the architecture is not chosen to efficiently encapsulate memory. We attribute this reasonable performance to the all-to-all communication between neurons and the fact that the FCNN does have access to the full driving trajectory. Essentially, it needs to learn causality on its own (but is of course less efficient than a dedicated architecture like LSTM, e.g. in terms of the total number of weights being much larger for the FCNN). See Appendix C. 1 for details related to the layout of the FCNN.

\section{Application in experiments}

We briefly indicate how our method would be applied to learn and predict quantum many-body dynamics in actual experiments. 

Obtaining the time evolution of any expectation value in an experiment involves several steps. For each time point $t$, the experiment has to be run $n$ times under a fixed driving trajectory until that point, when the measurement is performed. This will yield the expectation value up to an error $\sim 1/\sqrt{n}$ set by the projection noise. Afterwards, data has to be obtained for other time points and other driving trajectories. Overall, this requires $n N_t N$ runs, where $N$ is the number of different driving trajectories (training samples) and $N_t$ the number of time points. The number of runs is substantial, e.g. on the order of $10^{10}$ if we assume $N\sim 10^4$, $N_t\sim 10^2$ and $n\sim 10^4$ (for a 1 per cent projection noise error). Still, in platforms where a single run takes on the order of only a few microseconds, like superconducting qubits, this becomes feasible. The response of correlated quantum materials under fast high-field light pulses may be an even more favourable area of application. This is due to the shorter time scales involved and due to the parallelism stemming from the simultaneous observation of many copies of the system (many unit cells of the crystal providing a contribution to the overall signal which is directly a measure of the expectation value rather than a single projected measurement value). In general, multiple commuting observables can be measured in parallel, only non-commuting observables require separate runs.  In the future, it will be an interesting challenge to explore how much a NN can deal with more noisy measurement data, reducing the need to accumulate statistics. Some efficiency improvements are possible, e.g. one might allow for a larger statistical error (smaller $n$) if the time points are closely spaced, because the NN will then effectively try to interpolate smoothly through the noisy observations. One might also use tricks from the machine learning toolbox, like pre-training on simulated data (with a model that approximates the experiment at least roughly), which will condition the NN for more efficient learning on the actual experimental results.

\section{Conclusion\label{conclusion and discussion}}

In this work, we have introduced an approach that exploits the power of deep learning for predicting the dynamics of non-equilibrium quantum many-body systems.  The training is based on monitoring the expectation values of a subset of observables of interest under random driving trajectories.   We demonstrated that the trained RNN is able to produce accurate predictions for driving trajectories entirely different than those observed during training. Moreover, it is able to extrapolate to time spans and system sizes larger than those encountered during training.


Our scheme is model-free, i.e. learning requires no knowledge of the underlying quantum state or the correct evolution equations. In the future, this feature will allow for applications on actual experimental data, generated from a system that might even be open, noisy, inhomogeneous, or non-Markovian. This is in contrast to approaches which rely on precise knowledge of the underlying Hamiltonian, like exact Schr\"odinger evolution, tensor network wave functions, or recent methods based on NN representations of quantum states. Another independent benefit is the significant speedup that could be used, e.g., for pulse engineering via gradient-descent techniques or the exploration of dynamical phase diagrams. More advanced, feedback-based control strategies for quantum many-body systems might eventually be discovered from scratch in an efficient manner by combining the method introduced here with deep reinforcement learning.



\section*{Data Availability}
All data supporting the findings of this study are available within the article and supplementary information or from the first author upon request.

\section*{Code Availability}
The code used to generate the results of this study are available from the first author on request.

\begin{acknowledgments}
CNB thanks sponsorship from the Yangyang Development Fund, as well as support from a Shanghai talent program and from the Shanghai Municipal Science and Technology Major Project (Grant No. 2019SHZDZX01).

\end{acknowledgments}


\appendix
\section{Brief review of Recurrent Neural Networks}

Recurrent neural networks (RNNs) are built of a chain of repeating modules of neural networks (NNs). Such a NN introduces a feedback loop such that the output of the NN at the current time depends on the current input ($x_t$), called the external input, and also on the perceived information from the past, called the hidden input ($h_{t-1}$) \cite{nielsen2015neural}. Such a NN is able to record the history for -- in principle -- arbitrary long times, since weights are not time-dependent and therefore the number of trainable parameters does not grow with the time interval. For training such a NN, the gradient of the cost function needs to be backpropagated from the output towards the input layer, as in feedforward NNs, and also along backward along the time axis.

However, RNNs are prone to run into a fundamental problem, the ``vanishing/exploding gradient problem'', i.\,e., that the size of the gradient decreases (or sometimes increases) exponentially during backpropagation. In principle, this problem can also occur in traditional feedforward NNs, especially if they are deep. However, this effect is typically much stronger for RNNs since the time series can get very long. This seriously limited the trainability of early RNN architectures, which were not capable of capturing long-term dependencies. This problem led to the development of RNNs with cleverly designed gated units (controlling memory access) to avoid the exponential growth or vanishing of the gradient, and therefore permitting to train RNNs that capture both long-term dependencies. The first such architecture is called long short-term memory (LSTM), developed by Hochreiter and Schmidhuber in the late 90s \cite{hochreiter1735long}. 
\begin{figure}{}
	\centering
	\includegraphics[width=1.0\linewidth]{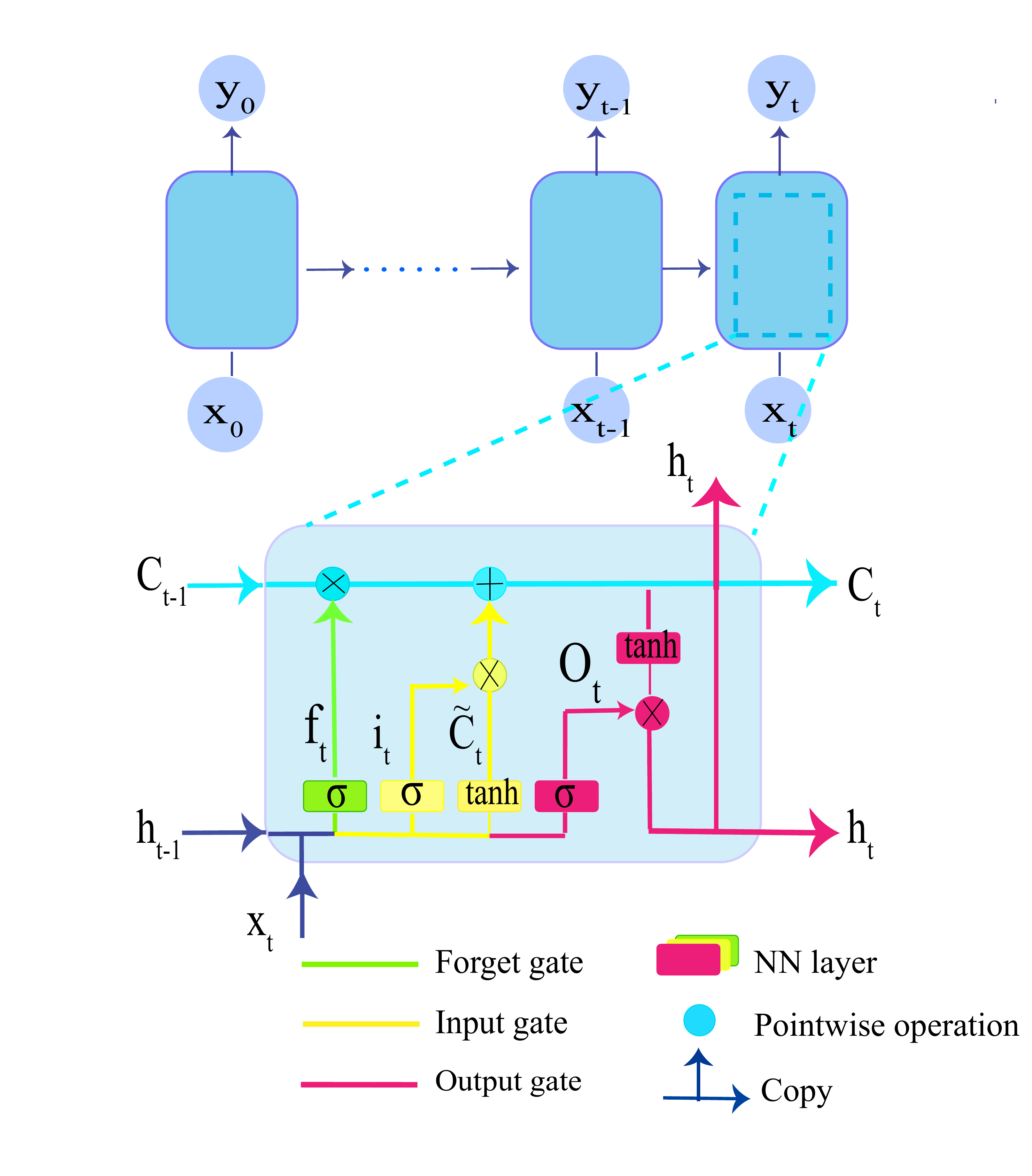}
	\caption{An LSTM-NN structure made of a chain of repeating modules, where each module includes three gates.}
	\label{figs1}
\end{figure}
As an RNN architecture, standard LSTM is also built of a chain of repeating modules, as is shown in Fig. \ref{figs1}, where the repeating modules have a more complicated structure than in a simple recurrent  network. Each module includes three gates, where each gate is composed out of a sigmoid NN layer, together with the point-wise multiplication on top of it. Next, we explain step by step how these three gates together control how the memory needs to be accessed. We label weights $w$ and biases $b$ by subscripts according to the name of the corresponding layer.

\begin{itemize}
	\item{Forget gate layer: this gate uses the hidden state $h_{t-1}$ from the previous time step and the external input $x_t$ at a particular time step $t$ (with the bias $b_f$ and the weight $w_f$) to decide whether to keep the memory, or to discard the information that is of less importance, applying a sigmoid activation.
		\begin{equation}
			f_{t}=\sigma\left(W_{f} \cdot\left[h_{t-1}, x_{t}\right]+b_{f}\right).
		\end{equation}
		$\sigma$ denotes sigmoid function and the dot stands for matrix multiplication. Eventually, the output of the forget gate is multiplied with the module state ($C_t$).}
	\item{Input gate layer: the operation of this gate is a three-step process,
		\begin{itemize}
			\item{first a sigmoid layer decides which data should be stored (very similar to the forget gate) 
				\begin{equation}
					i_{t} =\sigma\left(W_{i} \cdot\left[h_{t-1}, x_{t}\right]+b_{i}\right).
				\end{equation}}
			\item{hidden state and current input also will be passed into the tanh function to push values between -1 and 1 to regulate the NN and stored in $\tilde {C}_t$.
				\begin{equation}
					\tilde{C}_{t}=\tanh \left(W_{C} \cdot\left[h_{t-1}, x_{t}\right]+b_{C}\right).
				\end{equation}}
			\item{ the outcome of the two previous steps will be combined via multiplication operation and then this information is added to the module state ($f_{t} * C_{t-1}$).
				\begin{equation}
					C_{t}=f_{t} * C_{t-1}+i_{t} * \tilde{C}_{t}.
				\end{equation}}
		\end{itemize}
		Here the $*$ denotes element-wise multiplication.}
	\item{Output gate layer: the operation of this gate which decides the value of the next hidden input can be decomposed into two steps,
		\begin{itemize}
			\item{run a sigmoid layer on the previous hidden state and the current input, which decides what parts of the module state are going to be carried
				\begin{equation}
					\begin{array}{l}
						O_{t}=\sigma\left(W_{o}\left[h_{t-1}, x_{t}\right]+b_{o}\right). \\
					\end{array}
				\end{equation}}
			\item{passing the module state through tanh to squish the values to be between -1 and 1, and finally multiply it by the output of the sigmoid gate so that we only pass to the next module some selected parts
				\begin{equation}
					h_{t}=O_{t} * \tanh \left(C_{t}\right).
				\end{equation}}
		\end{itemize}}
\end{itemize}

\section{Generating random driving trajectories}

In this section we explain how we generate our random samples for training and also for evaluating the NN.


\subsection{Sampling driving trajectories from a Gaussian random process for training the neural network}
There are different methods to generate  Gaussian random functions \cite{liu2019advances}. We will explain in detail the one we use. Discretizing the time interval of interest according to some step $\Delta t$, we define a vector $\boldsymbol{d}=(D(0),D(\Delta t),D(2\Delta t),...)^T$ that contains the trajectory of the time-dependent parameter. We build the correlation matrix $C$ with elements $C_{nm}=\langle d_n d_m\rangle=c_0\exp[-(n-m)^2\Delta t^2/2\sigma^2)]$, where in this case we assumed a Gaussian correlation function with a correlation time $\sigma$ (though other functional forms could be used). Being real and symmetric, this can be diagonalized as $C=Q\Lambda Q^T$, where $\Lambda$ is a diagonal matrix containing the eigenvalues and $Q$ is an orthogonal matrix. Hence, we can generate the random parameter trajectory as $\boldsymbol{d}=Q\sqrt{\Lambda}\boldsymbol{x}$, where the components of $\boldsymbol{x}$ are independent random variables drawn from the unit-width normal distribution ($\langle x_n\rangle=0$ and $\langle x_n x_m\rangle=\delta_{nm}$), which can be easily generated. Note that we could sample using Fourier series with random coefficients, but this would automatically generate periodic functions, and thus introduce spurious correlations between the values of the drive at early and late times. That is why we avoided that method.

In order to make sure that our generated random functions are characteristic for a wide range of arbitrary time-dependent functions, we also choose for any training trajectory the correlation amplitude $c_0$ and time $\sigma$ randomly. The former is chosen uniformly from $c_0\in[0,4]$, corresponding to a maximum value of the magnetic field $|B|$ around 5, and the latter uniformly from the wide interval $\sigma \in[1,9]$. Technically, this means we are sampling from a mixture of Gaussian processes.

\subsection{Generating random time-dependent functions for evaluating the network}

As we explained in the main text, in order to evaluate the performance of the NN, and to illustrate that it succeeds in predicting the dynamics for different functional forms of the driving trajectories, we test the NN on random periodic driving fields and different sorts of quenches. As for the periodic drivings, we generate 1000 functions of the form $B(t)=A sin(\omega t) $ with random amplitudes $A$ and frequencies $\omega$ chosen uniformly from intervals $[-3,3]$ and $[0.1, 4]$, respectively. As for quenches, we generate 1000 step functions where the heights of the steps are chosen randomly from the interval $[-3,3]$ with the quench also happening at random times.

\section{Neural network layout}

In this section, we present the details regarding the layout of all the architectures that we have investigated. We have specified and trained all these different architectures with Keras \cite{chollet2015keras}, a deep-learning framework written for Python.

\subsection{Fully connected and LSTM neural network}

In Table. \ref{table_1}, we present the layout of our deep fully-connected neural network (FCNN) and the LSTM-NN. For the FCNN, the activation function for all layers except the last layer is the rectified linear activation function (``$\mathrm{ReLU}$''). For the last layer the activation function is ``linear''. As optimizer we always use ``adam''.

\begin{table}[!h]
\scalebox{0.9}{
	\begin{tabular}{|l|l|l|l|l|l|}
		\hline NN architecture & L & N & Input size & Output size \\
		\hline FCNN & 6 & 800&n+3 &9(l-1)+3 \\
		\hline LSTM & 4 & 500&1+1+3&9(l-1)+3 \\
		\hline
	\end{tabular}}
	\caption{
		The layout of the deep fully-connected neural network (FCNN) and of the LSTM-NN. The $L$, $N$, and $n$ stands for number of hidden layers, neurons per layer, and time steps. The $l$ denotes the largest distance between spins on a ring. The expression $9(l-1)+3$ counts the total number of all the observables that we choose for training throughout the examples discussed in the main text, containing all first and second-order moments (correlators) of spin operators. The number 3 is related to the number of parameters that identify the initial state. Our initial product state is identified by the 3 expectation values of first-order moments of the spin operators.
	}
	\label{table_1}
\end{table}

The most important difference between these two architectures is that the LSTM receives as input only the  information for a single time step and outputs the predicted observables for that time step (as can be seen in the number of input and output neurons), eventually working its way sequentially through the whole time interval. Note that because the LSTM cells provide memory, all the information from previous time steps is still available. In contrast, the FCNN receives as input the full driving trajectory at once and also outputs the full time-dependent evolution of the observables throughout the whole time interval at once. For the LSTM, besides the driving field, we also feed in the current value of the time variable, which seems to help to enhance the accuracy of predictions (this is reasonable, since time-translation invariance is broken by the preparation of the initial state at time 0). In addition, both the FCNN and the LSTM receive as input information about the initial quantum state, in the form of the expectation values of the spin components (which is sufficient as we consider only product states as initial states).


\subsection{Convolutional neural network}

\begin{figure}
	\centering
	\includegraphics[width=1\linewidth]{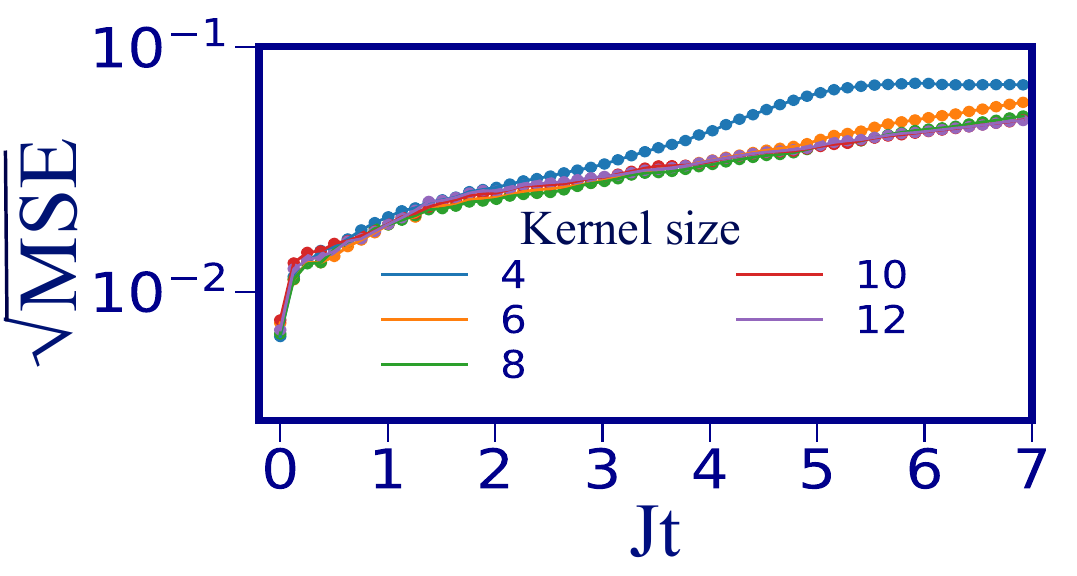}
	\caption{The performance of a temporal 1D-CNN for different kernel sizes on the TFI model driven with  Gaussian driving random fields, ($M=15$). The $\sqrt{\textrm{MSE}}$ versus time averaged over all the selected observables and over 1000 realizations of Gaussian random driving fields.}
	\label{figS3}
\end{figure}

In this section, we present the layout of the temporal 1D-CNN that we apply for predicting the dynamics of the TFI model in the main text. We use the ``$\mathrm{ReLU}$'' activation for all layers. Our NN contains 13 layers, with filter sizes (50,80,120,150,150,200,300,300,200,150,80,50,9(l-1)+3), respectively. Here, $l$ denotes the largest distance between spins in the ring. The kernel sizes for all layers are chosen as (8,8,8,8,8,8,8,8,6,6,6,4,4), respectively. We set the padding parameter as ``padding=causal'', which means the output at time $t$ only depends on the previous time steps; therefore, the NN has a notion of causality. In Fig. \ref{figS3}, we also compare the performance of this architecture for different kernel sizes. As it is evident, a larger kernel size leads to a better performance at longer times. However, above kernel size=10, the performance almost saturates for the 56 time steps that we considered here.

\section{Computational resources}

\begin{figure}[!h]
	\centering
	\includegraphics[width=1\linewidth]{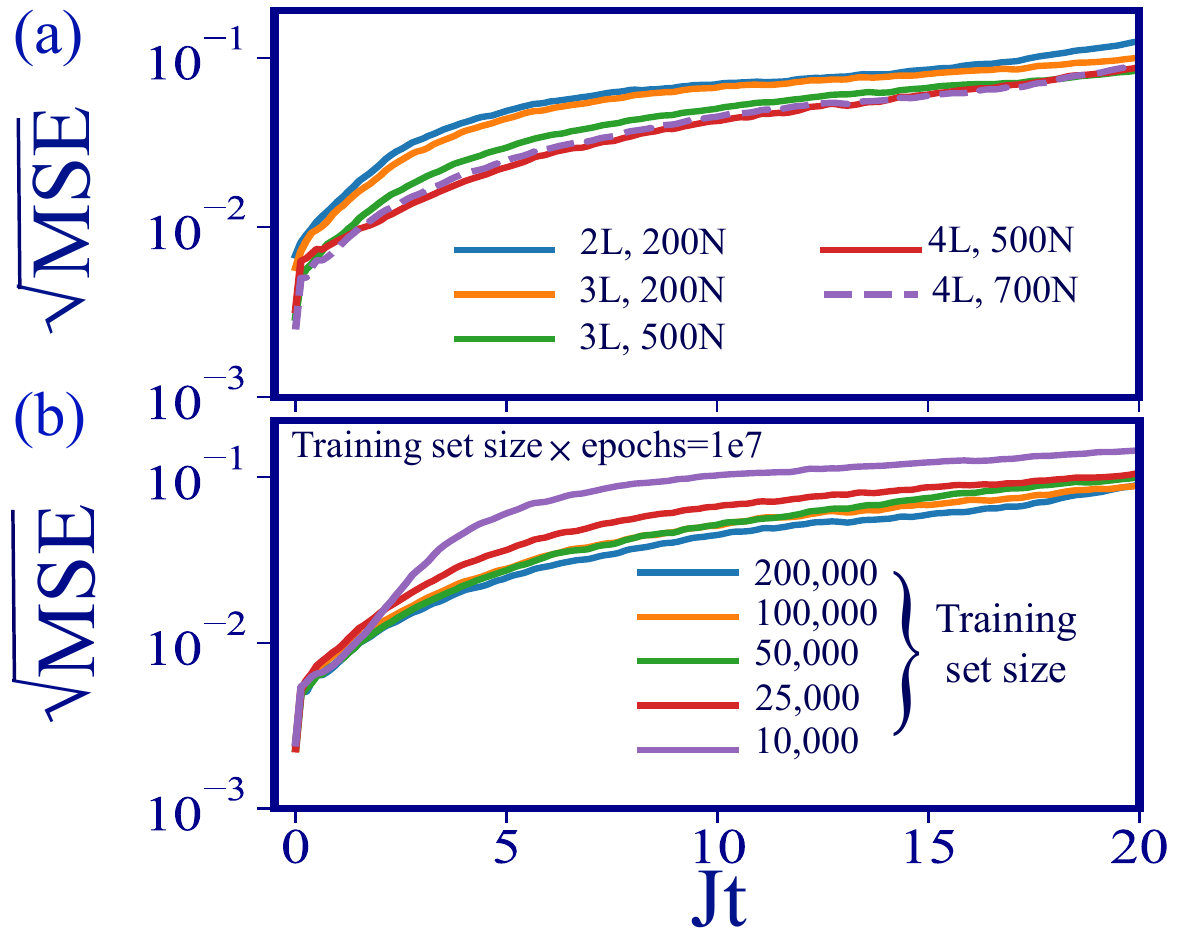}
	\caption{
		Role of the LSTM-NN size and training set size on the performance of LSTM-NN.
		(a) The $\sqrt{\textrm{MSE}}$ versus time for different NN size. $L$ and $N$ stand for number of layers and number of neurons, respectively.
		(b) The $\sqrt{\textrm{MSE}}$ versus time for different training set size.
	}
	\label{figS2}
\end{figure}

In this section, we provide details related to the computational resources required for training. In Fig. \ref{figS2}(a), we show how the precision of the LSTM-NN in predicting the dynamics of the TFI model with $M=7$ is related to the size of the NN.

We also discuss how for the same model the performance of the NN is related to the training set size. Fig. \ref{figS2}(b) shows how error scales versus time for different training set size when the NN is trained on a fixed number of instances. We define number of instances as training set size multiplied by the number of epochs. The number of epochs denotes the number of times that the NN goes through the entire training data set.  As can be appreciated, 50,000 samples already provides a favourable performance. We observed more or less the same scaling behaviour (in terms of the size of the NN and the training set size) for the rest of the models and different system sizes that we checked out.

We generated the training samples on CPU nodes, and trained the NNs on GPU nodes of a high-performance computing cluster. In Table \ref{table_3}, we provide details related to the required resources for different models and system sizes that we have checked out, given 50,000 samples are provided for training.

\begin{table}[h]
\scalebox{0.75}{
	\begin{tabular}{|l|l|l|l|l|l|}
		\hline Architecture & \# Samples&Batch size & Time/Epoch & \#Epochs \\
		\hline FCNN & 200,000 & 1000 & 29 second & 30 \\
		\hline CNN  & 200,000 & 1000 & 26 second & 100 \\
		\hline LSTM & 50,000 & 1000 & 21 second & 200 \\
		\hline
	\end{tabular}}
	\caption{Resources required for training the different architectures for TFI model with $M=7$ to reach their best performance. Batch size denotes size of the dataset over which the gradient is calculated and weights get updated.}
	\label{table_3}
\end{table}

\section{Performance of the LSTM network trained on first-order moments of spin operators}
\begin{figure}
	\centering
	\includegraphics[width=1\linewidth]{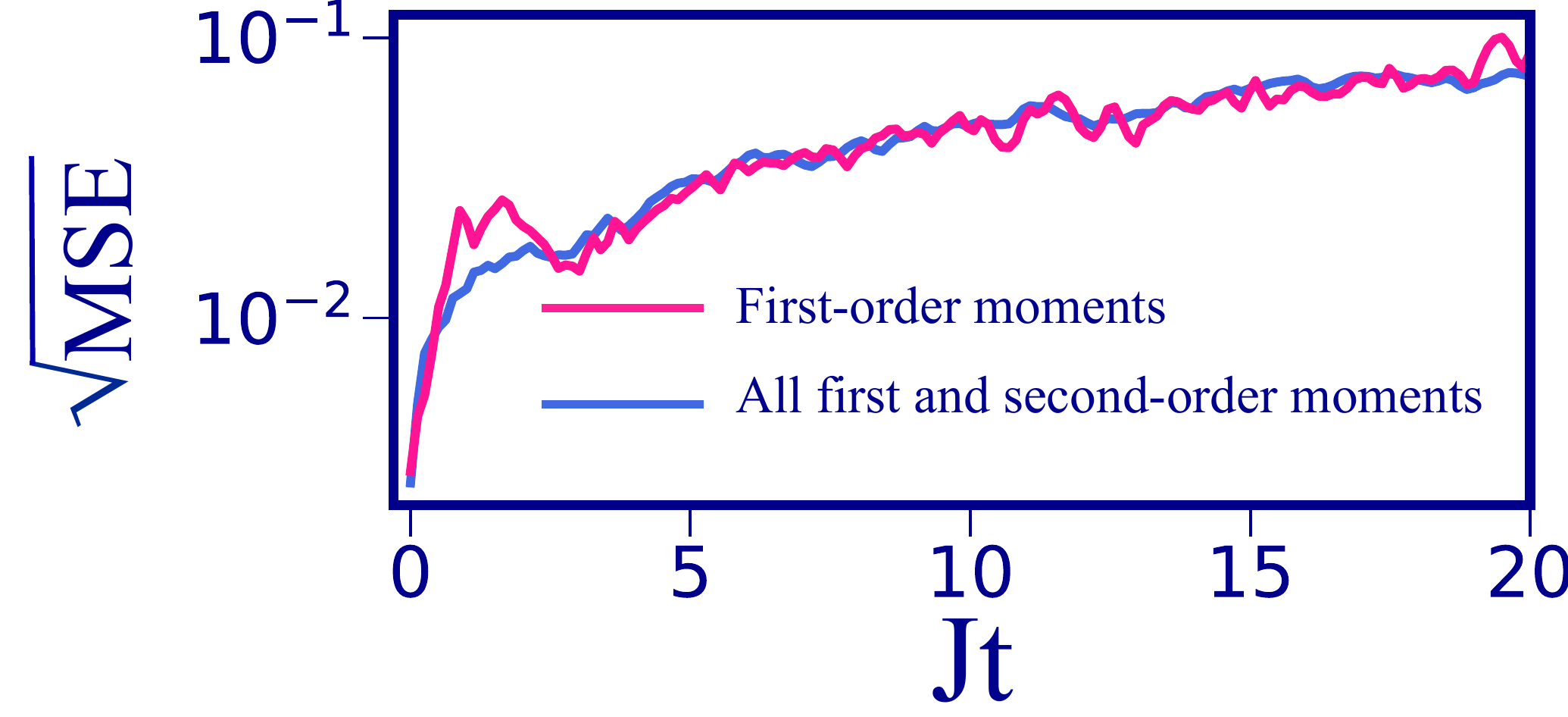}
	\caption{Comparing the performance of the NN for the  case where it is trained just over first-order moments of spin operators, vs. the case where it is trained over all first and second-order moments of spin operators. The $\sqrt{\textrm{MSE}}$ versus time averaged over all the selected observables and over 1000 realizations of  Gaussian random  driving fields.}
	\label{figS4}
\end{figure}

In this work we always trained the NN on all first and second-order moments of spin operators.
However, this choice was for illustration purposes only. We demonstrate here that one does not need to train for all these observables to get the NN to learn the dynamics properly.
In Fig. \ref{figS4}, we show the error versus time, comparing the two cases where we either trained the NN  on just first-order moments of spin operators or over all first and second-order moments of spin operators. These results are for a TFI model with $M=11$. This figure verifies that  one can successfully train the NN on fewer observables. 
\section{ Power of the LSTM network in implicit construction of higher-order correlators}
\begin{figure}
	\centering
	\includegraphics[width=1\linewidth]{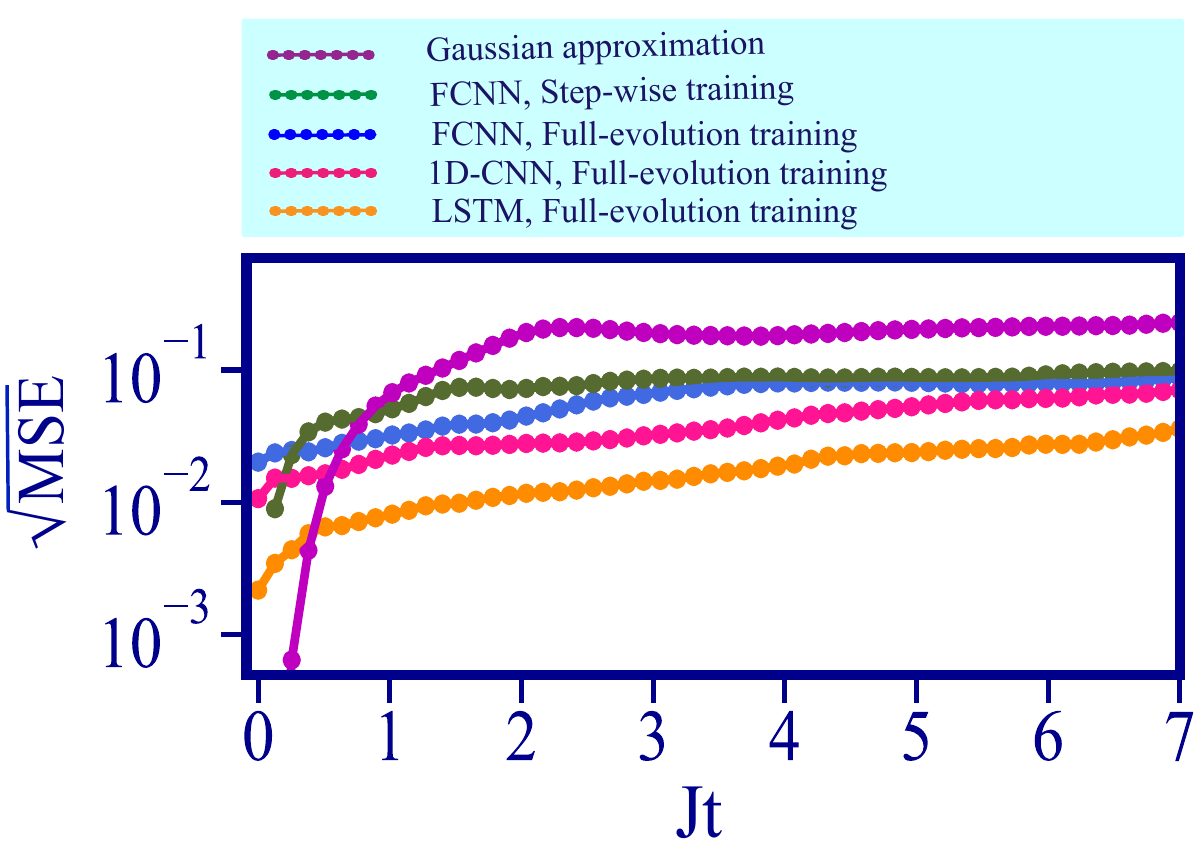}
	\caption{Comparing the performance of different NN architectures against Gaussian approximation.}
	\label{figS5}
\end{figure}

In the main text, to illustrate the power of the LSTM-NN in building an implicit construction for higher-order correlators, we compared the performance of the LSTM-NN against truncated equations of motion within the Gaussian approximation for a typical driving field. To demonstrate this is generally true even for other architectures, here we compare them for different architectures on 1000 realizations of different sorts of quenches. However, we emphasize that our goal in presenting this plot is not at all to compare NN with a very naive approximation. Instead, we aim to illustrate that our NNs are able to implicitly construct higher-order correlators. As visible from Fig. \ref{figS5}, all architectures outperform the Gaussian approximation. It is surprising that above some time, an FCNN that is trained in a step-wise manner and has no chance for constructing an implicit representation of higher-order correlators still outperforms the Gaussian approximation. We imagine that this is due to the fact that  the dynamics does not explore completely arbitrary states in this case and therefore still the NN can approximately infer some higher-order correlators from the observables which it is supplied with as input at any given time. 

\begin{figure}
	\centering
	\includegraphics[width=1\linewidth]{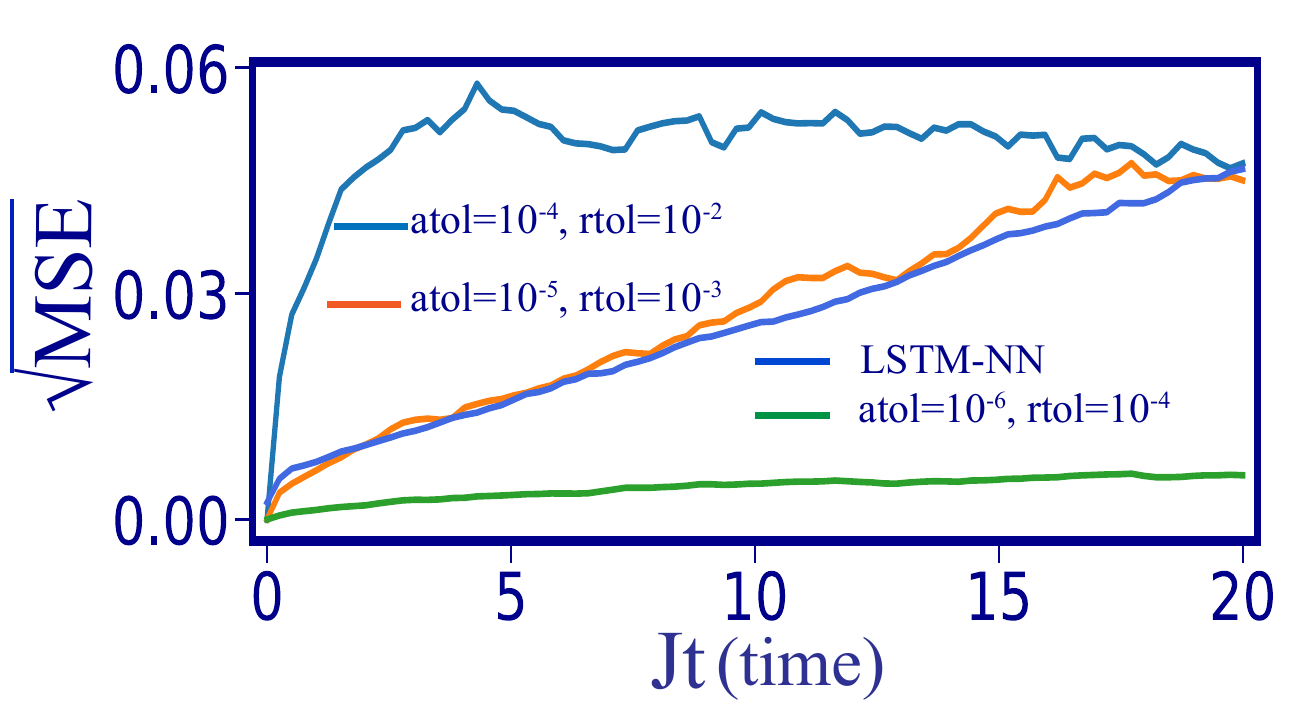}
	\caption{ Comparing the error in predicting the dynamics of the observables by the LSTM-NN vs. the error in numerical integration for different relative  and  absolute tolerances. The shown results are related to the driven TFI model with $M=15$ and we averaged over 100 realizations of random driving fields.
	}
	\label{figS6}
\end{figure}
\section{Accuracy in solving the Schr\"odinger equation}
As we indicated in the main text, for solving the Schr\"odinger equation we used Qutip \cite{johansson2012qutip}, which uses scipy 
and ``zvode'' method for the integration. To compare the runtime of the integrator with the NN,  we relaxed the accuracy of the integrator by increasing the relative  and  absolute tolerance denoted by ``rtol" and ``atol", respectively. This results in a larger integration step. In Fig. ~\ref{figS6}, we compare the error in predicting the dynamics of the observables by the NN vs. the error in numerical integration with different relative  and  absolute tolerances. Note that we trained the NN on the dynamics achieved  by our integrator with a high accuracy namely small stepsize. We define  the $\textrm{MSE}$ as the quadratic deviation of  the prediction gained by the NN and the dynamics achieved by our integrator with a relaxed accuracy with the dynamics achieved by our integrator with high accuracy.  As can be seen, setting the ``atol=$10^{-5}$" and the ``rtol=$10^{-3}$",  the error in numerical integration is comparable with the error observed by the NN. Therefore, as runtime for the integrator, we count the runtime correspond to this accuracy.

\section{Power of the LSTM network in predicting the dynamics of a non-integrable model}
\begin{figure}[t]
	\centering
	\includegraphics[width=1\linewidth]{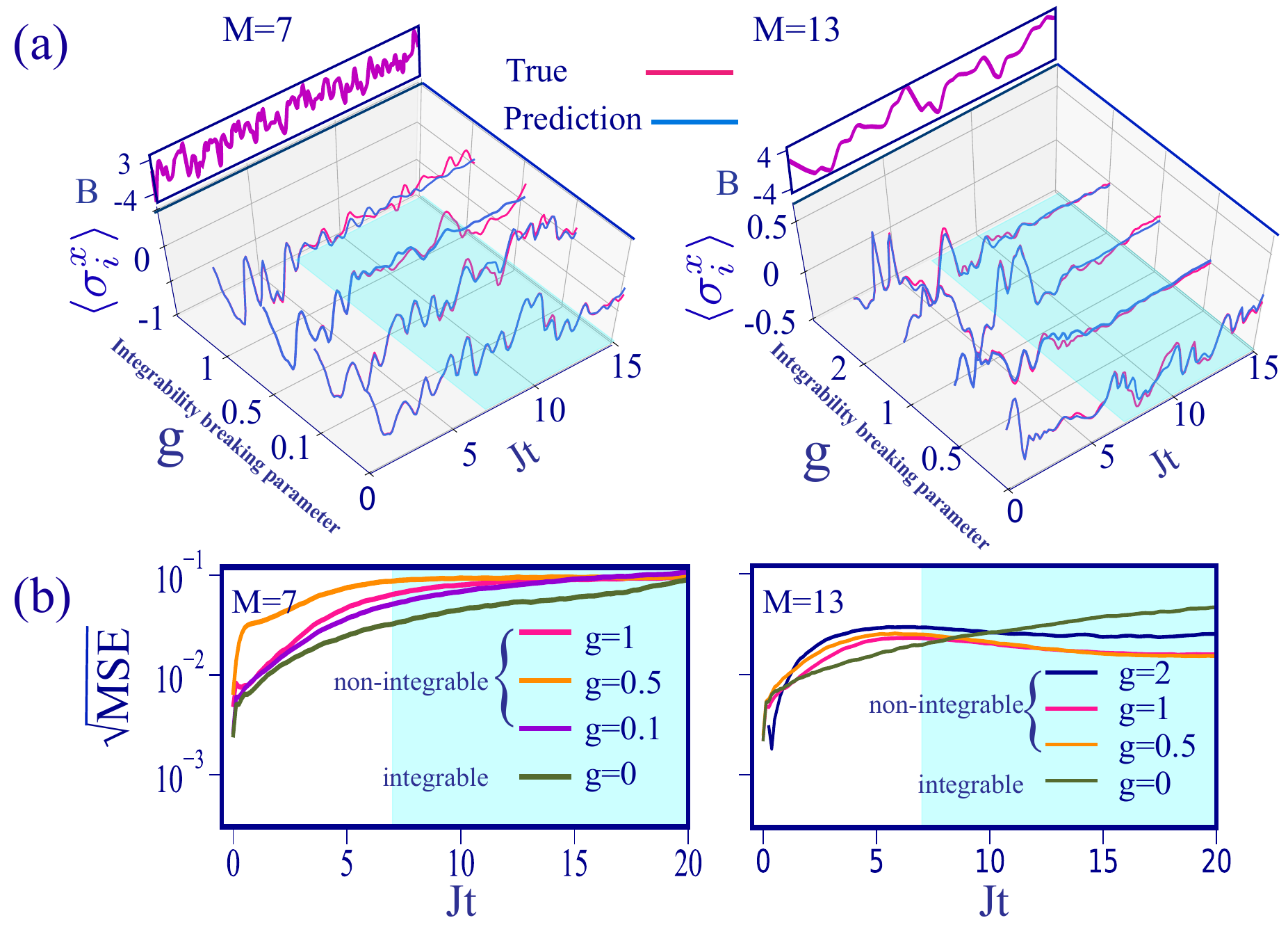}
	\caption{
		Comparing the power of the LSTM-NN in predicting the dynamics of quantum-integrable models versus quantum non-integrable models.
		(a) Comparing the performance of the LSTM-NN in predicting $\sigma_{i}^{x}$ against the true evolution for different values of $g$ under the shown driving fields.
		(b) The $\sqrt{\textrm{MSE}}$ versus time in predicting all the first and second-order moments of spin operators for different values of $g$ where the spin-ring is driven by  Gaussian random transverse fields. For each value of $g$, the NN is evaluated on 1000 realizations for system sizes $M=7$ and $M=13$. The blue highlighted region shows the interval that the NN has not been trained on.
	}
	\label{figS7}
\end{figure}

In Fig.~ \ref{figS7}, we show the performance of the LSTM-NN in predicting the dynamics of the model (5) presented in Sec. V-E in the main text, subject to Gaussian random transverse fields (as opposed to quenches). 
We show the performance for a single instance (Fig.~ \ref{figS7}(a)) and  over 1000 random realizations for different  values of $g$ (Fig.~ \ref{figS7} (b)). It seems that the NN is less accurate in predicting the dynamics when $g \neq 0$ in comparison with  $g=0$. This is more evident for $M=7$. As we indicated in the main text, we observed for the small system sizes that the NN has less power for $g \neq 0$ to predict the finite size effects. In such cases the NN has a tendency to predict trajectories close to the infinite-system-size limit. 

For $M=13$, the errors for non-integrable models become smaller for longer times. However, this does not imply that the NN has  a  better performance. As it is visible in panel (b), the reason is that at longer times, the dynamics for non-integrable cases relaxes to an almost fixed value.

\section{Heisenberg model}

\begin{figure}[ht]
	\centering
	\includegraphics[width=1\linewidth]{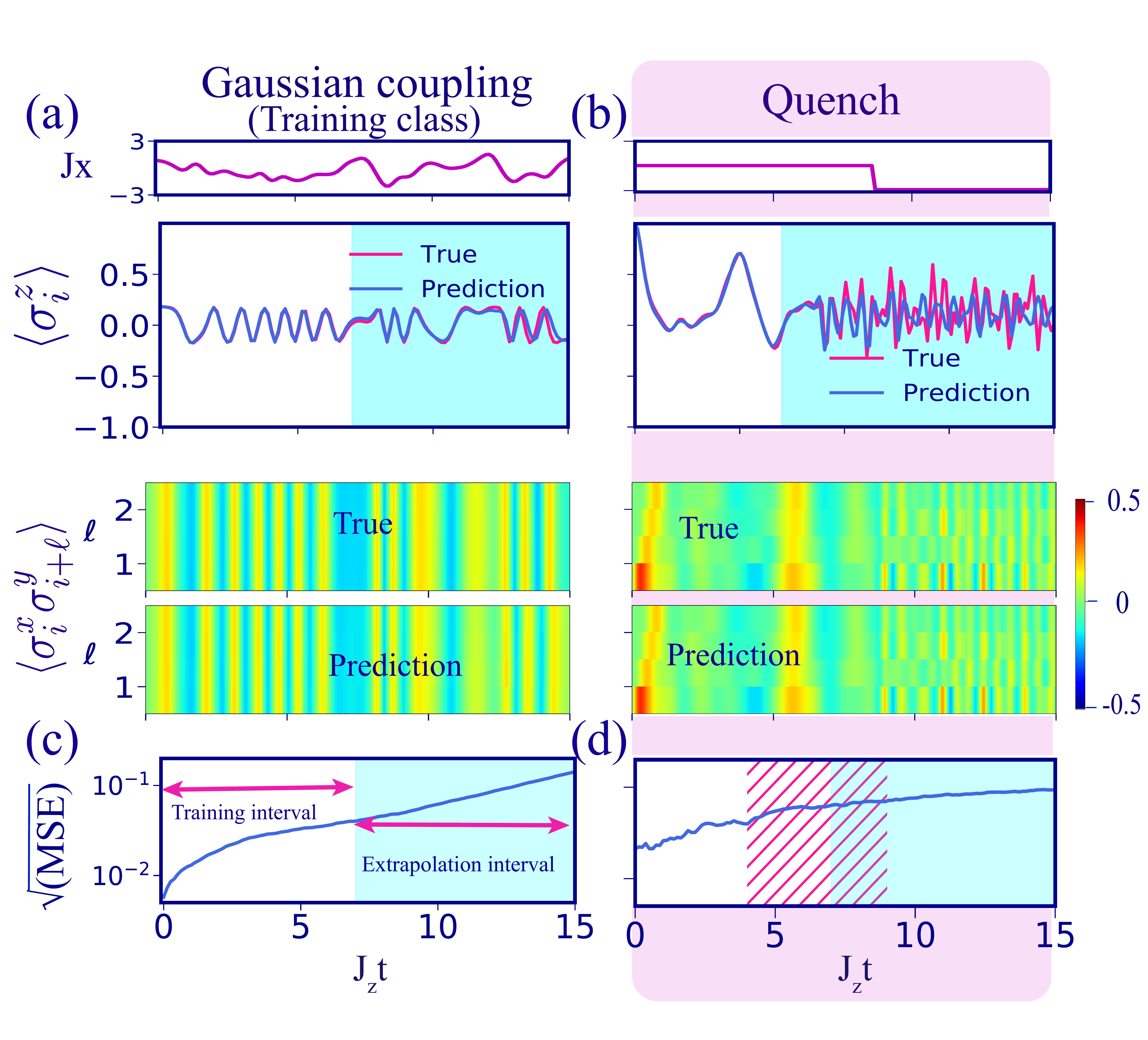}
	\caption{
		The power of the LSTM-NN in predicting the dynamics of a Heisenberg spin ring under time-dependent exchange couplings with $M=9$. The NN is trained on  Gaussian random exchange couplings in $J_z t \in[0,7]$ and then is evaluated on a new set of  Gaussian random exchange couplings as well as different sorts of quenches in the time window $J_z t \in[0,15]$. Comparing the real and predicted evolution of $\langle \sigma^x_i \rangle$ and $\langle \sigma^{x}_{i}\sigma^{y}_{i+\ell}\rangle$ for different values of $\ell$ under the shown (a)  Gaussian random coupling (b) quench (in direction of $x$) at the top of each panel. The highlighted blue regions denote the time window that the NN has not been trained on. The highlighted light pink column remarks that NN is evaluated on the class of couplings that the NN has never seen. The $\sqrt{\textrm{MSE}}$ in predicting all first and second-order moments of spin operators where the spin ring is driven by (c) 1000  Gaussian random exchange couplings, (d) 1000 random quenches in direction of $x$. The striped area shows the region in which the time of the jump has been randomly chosen.
	}
	\label{figS8}
\end{figure}

In this section, we illustrate the power of the LSTM-NN in predicting the dynamics of the 1D-Heisenberg model. We take advantage of this example to illustrate the power of our NN for the models where we drive the coupling constants (instead of the fields). 
We train the NN on the dynamics of following  Hamiltonian
\begin{equation}
	\mathcal{H}_{\mathrm{Heis}}=J_{x} (t)\sum_{i=1}^M \sigma_{i}^{x} \sigma_{i+1}^{x}+J_{y}  \sum_{i=1}^M\sigma_{i}^{y} \sigma_{i+1}^{y}+J_{z} \sum_{i=1}^M \sigma_{i}^{z} \sigma_{i+1}^{z},
\end{equation}
where $J_x$ is a homogeneous Gaussian random function and we set $J_z=1$ and $J_y=1$.

We train the NN on system size $M=9$ up to $Jt=7$ on  Gaussian random exchange couplings. Eventually, we evaluate the NN in predicting the dynamics for the same system size up to $Jt=15$ for a new set of random exchange couplings as well as quenches on couplings. In Fig. \ref{figS8} (a), we show the performance of our LSTM-NN in predicting the dynamics under the  shown  Gaussian random exchange coupling in direction of $ x $ at the top of the panel. We compare the true and predicted evolution of $\langle \sigma^z_i \rangle$ and $\langle \sigma^{x}_{i}\sigma^{y}_{i+\ell} \rangle$ for different values of $\ell$. As can be seen, the NN is able to predict the dynamics very well for the time window that it has been trained on as well as extrapolating the dynamics to a longer time window which is highlighted in light blue. In Fig. \ref{figS8} (b), we show the performance of the NN on the class of exchange couplings that it has never seen, namely quench in $x$  as shown at the top of the panel.  As can be appreciated, the NN is able to predict the shown first and second-order moments for the time-window that it has been trained on, and  for longer times. 

To illustrate that comparable results can be achieved for a wide variety of exchange couplings, we show in Fig.  \ref{figS8}  (c) and (d) the $\sqrt{\textrm{MSE}}$ -- the square root of quadratic deviation between the true dynamics and the predictions obtained from NN -- on predicting the subset of observable of interest. For each class of exchange couplings, we averaged on 1000 realizations. These plots demonstrate the power of our LSTM-NN in predicting the dynamics of this model for the time interval that it has been trained on and beyond that for a big variety of samples. For both classes of exchange couplings, the spins are initially prepared in a randomly chosen translationally invariant product state.

Checking out the performance of our NN  for a few more system sizes, we observed sometimes the NN is less accurate in predicting finite-size effects at longer times. For these cases the NN has a tendency to predict trajectories close to the infinite-system-size limit, rather than the given system size that it has been trained on. The reason for this remains open and needs to be explored more as part of future work.


\begin{thebibliography}{52}
\providecommand{\natexlab}[1]{#1}
\providecommand{\url}[1]{\texttt{#1}}
\expandafter\ifx\csname urlstyle\endcsname\relax
  \providecommand{\doi}[1]{doi: #1}\else
  \providecommand{\doi}{doi: \begingroup \urlstyle{rm}\Url}\fi

\bibitem[Anderlini et~al.(2007)Anderlini, Lee, Brown, Sebby-Strabley, Phillips,
  and Porto]{anderlini2007controlled}
Marco Anderlini, Patricia~J Lee, Benjamin~L Brown, Jennifer Sebby-Strabley,
  William~D Phillips, and James~V Porto.
\newblock Controlled exchange interaction between pairs of neutral atoms in an
  optical lattice.
\newblock \emph{Nature}, 448\penalty0 (7152):\penalty0 452--456, 2007.
\newblock URL \url{https://doi.org/10.1038/nature06011}.

\bibitem[Beach et~al.(2018)Beach, Golubeva, and Melko]{beach2018machine}
Matthew~JS Beach, Anna Golubeva, and Roger~G Melko.
\newblock Machine learning vortices at the kosterlitz-thouless transition.
\newblock \emph{Physical Review B}, 97\penalty0 (4):\penalty0 045207, 2018.
\newblock URL \url{https://doi.org/10.1103/PhysRevB.97.045207}.

\bibitem[Calabrese and Cardy(2005)]{Calabrese_2005}
Pasquale Calabrese and John Cardy.
\newblock Evolution of entanglement entropy in one-dimensional systems.
\newblock \emph{Journal of Statistical Mechanics: Theory and Experiment},
  2005\penalty0 (04):\penalty0 P04010, apr 2005.
\newblock \doi{10.1088/1742-5468/2005/04/p04010}.
\newblock URL \url{https://doi.org/10.1088%2F1742-5468%2F2005%2F04%2Fp04010}.

\bibitem[Carleo and Troyer(2017)]{carleo2017solving}
Giuseppe Carleo and Matthias Troyer.
\newblock Solving the quantum many-body problem with artificial neural
  networks.
\newblock \emph{Science}, 355\penalty0 (6325):\penalty0 602--606, 2017.
\newblock URL \url{https://www.science.org/doi/10.1126/science.aag2302}.

\bibitem[Carrasquilla and Melko(2017)]{carrasquilla2017machine}
Juan Carrasquilla and Roger~G Melko.
\newblock Machine learning phases of matter.
\newblock \emph{Nature Physics}, 13\penalty0 (5):\penalty0 431--434, 2017.
\newblock URL \url{https://doi.org/10.1038/nphys4035}.

\bibitem[Chakrabarti et~al.(2008)Chakrabarti, Dutta, and
  Sen]{chakrabarti2008quantum}
Bikas~K Chakrabarti, Amit Dutta, and Parongama Sen.
\newblock \emph{Quantum Ising phases and transitions in transverse Ising
  models}, volume~41.
\newblock Springer Science \& Business Media, 2008.
\newblock URL \url{https://doi.org/10.1007/978-3-642-33039-1}.

\bibitem[Cheneau et~al.(2012)Cheneau, Barmettler, Poletti, Endres, Schau{\ss},
  Fukuhara, Gross, Bloch, Kollath, and Kuhr]{cheneau2012light}
Marc Cheneau, Peter Barmettler, Dario Poletti, Manuel Endres, Peter Schau{\ss},
  Takeshi Fukuhara, Christian Gross, Immanuel Bloch, Corinna Kollath, and
  Stefan Kuhr.
\newblock Light-cone-like spreading of correlations in a quantum many-body
  system.
\newblock \emph{Nature}, 481\penalty0 (7382):\penalty0 484--487, 2012.
\newblock URL \url{https://doi.org/10.1038/nature10748}.

\bibitem[Choi et~al.(2020)Choi, Zhou, Knowles, Landig, Choi, and
  Lukin]{PhysRevX.10.031002}
Joonhee Choi, Hengyun Zhou, Helena~S. Knowles, Renate Landig, Soonwon Choi, and
  Mikhail~D. Lukin.
\newblock Robust dynamic hamiltonian engineering of many-body spin systems.
\newblock \emph{Phys. Rev. X}, 10:\penalty0 031002, Jul 2020.
\newblock \doi{10.1103/PhysRevX.10.031002}.
\newblock URL \url{https://link.aps.org/doi/10.1103/PhysRevX.10.031002}.

\bibitem[Chollet et~al.(2015)]{chollet2015keras}
Fran\c{c}ois Chollet et~al.
\newblock Keras.
\newblock \url{https://keras.io}, 2015.

\bibitem[Czischek et~al.(2018)Czischek, G\"arttner, and
  Gasenzer]{PhysRevB.98.024311}
Stefanie Czischek, Martin G\"arttner, and Thomas Gasenzer.
\newblock Quenches near ising quantum criticality as a challenge for artificial
  neural networks.
\newblock \emph{Phys. Rev. B}, 98:\penalty0 024311, Jul 2018.
\newblock \doi{10.1103/PhysRevB.98.024311}.
\newblock URL \url{https://link.aps.org/doi/10.1103/PhysRevB.98.024311}.

\bibitem[Eckardt(2017)]{RevModPhys.89.011004}
Andr\'e Eckardt.
\newblock Colloquium: Atomic quantum gases in periodically driven optical
  lattices.
\newblock \emph{Rev. Mod. Phys.}, 89:\penalty0 011004, Mar 2017.
\newblock \doi{10.1103/RevModPhys.89.011004}.
\newblock URL \url{https://link.aps.org/doi/10.1103/RevModPhys.89.011004}.

\bibitem[Flurin et~al.(2020)Flurin, Martin, Hacohen-Gourgy, and
  Siddiqi]{flurin2020using}
Emmanuel Flurin, Leigh~S Martin, Shay Hacohen-Gourgy, and Irfan Siddiqi.
\newblock Using a recurrent neural network to reconstruct quantum dynamics of a
  superconducting qubit from physical observations.
\newblock \emph{Physical Review X}, 10\penalty0 (1):\penalty0 011006, 2020.

\bibitem[Foroozandeh et~al.(2014)Foroozandeh, Adams, Meharry, Jeannerat,
  Nilsson, and Morris]{foroozandeh2014ultrahigh}
Mohammadali Foroozandeh, Ralph~W Adams, Nicola~J Meharry, Damien Jeannerat,
  Mathias Nilsson, and Gareth~A Morris.
\newblock Ultrahigh-resolution nmr spectroscopy.
\newblock \emph{Angewandte Chemie International Edition}, 53\penalty0
  (27):\penalty0 6990--6992, 2014.
\newblock URL \url{https://doi.org/10.1002/anie.201404111}.

\bibitem[F\"osel et~al.(2018)F\"osel, Tighineanu, Weiss, and
  Marquardt]{PhysRevX.8.031084}
Thomas F\"osel, Petru Tighineanu, Talitha Weiss, and Florian Marquardt.
\newblock Reinforcement learning with neural networks for quantum feedback.
\newblock \emph{Phys. Rev. X}, 8:\penalty0 031084, Sep 2018.
\newblock \doi{10.1103/PhysRevX.8.031084}.
\newblock URL \url{https://link.aps.org/doi/10.1103/PhysRevX.8.031084}.

\bibitem[Gao and Duan(2017)]{gao2017efficient}
Xun Gao and Lu-Ming Duan.
\newblock Efficient representation of quantum many-body states with deep neural
  networks.
\newblock \emph{Nature communications}, 8\penalty0 (1):\penalty0 1--6, 2017.
\newblock \doi{10.1038/s41467-017-00705-2}.
\newblock URL \url{https://doi.org/10.1038/s41467-017-00705-2}.

\bibitem[Glasser et~al.(2018)Glasser, Pancotti, August, Rodriguez, and
  Cirac]{Prx}
Ivan Glasser, Nicola Pancotti, Moritz August, Ivan~D. Rodriguez, and J.~Ignacio
  Cirac.
\newblock Neural-network quantum states, string-bond states, and chiral
  topological states.
\newblock \emph{Phys. Rev. X}, 8:\penalty0 011006, Jan 2018.
\newblock \doi{10.1103/PhysRevX.8.011006}.
\newblock URL \url{https://link.aps.org/doi/10.1103/PhysRevX.8.011006}.

\bibitem[Goodfellow et~al.(2016)Goodfellow, Bengio, Courville, and
  Bengio]{goodfellow2016deep}
Ian Goodfellow, Yoshua Bengio, Aaron Courville, and Yoshua Bengio.
\newblock \emph{Deep learning}, volume~1.
\newblock MIT press Cambridge, 2016.
\newblock URL \url{https://dl.acm.org/doi/book/10.5555/3086952}.

\bibitem[Graves et~al.(2013)Graves, Mohamed, and Hinton]{graves2013speech}
Alex Graves, Abdel-rahman Mohamed, and Geoffrey Hinton.
\newblock Speech recognition with deep recurrent neural networks.
\newblock In \emph{2013 IEEE international conference on acoustics, speech and
  signal processing}, pages 6645--6649. Ieee, 2013.
\newblock URL \url{https://doi.org/10.1109/ICASSP.2013.6638947}.

\bibitem[Greiner et~al.(2002)Greiner, Mandel, H{\"a}nsch, and
  Bloch]{greiner2002collapse}
Markus Greiner, Olaf Mandel, Theodor~W H{\"a}nsch, and Immanuel Bloch.
\newblock Collapse and revival of the matter wave field of a bose--einstein
  condensate.
\newblock \emph{Nature}, 419\penalty0 (6902):\penalty0 51--54, 2002.
\newblock URL \url{https://doi.org/10.1038/nature00968}.

\bibitem[Heyl(2012)]{heyl2012nonequilibrium}
Markus Philip~Ludwig Heyl.
\newblock \emph{Nonequilibrium phenomena in many-body quantum systems}.
\newblock PhD thesis, lmu, 2012.
\newblock URL \url{https://doi.org/10.5282/edoc.14583}.

\bibitem[Hochreiter and Schmidhuber(1997)]{hochreiter1735long}
Sepp Hochreiter and J{\"u}rgen Schmidhuber.
\newblock Long short-term memory.
\newblock \emph{Neural computation}, 9\penalty0 (8):\penalty0 1735--1780, 1997.
\newblock URL \url{https://doi.org/10.1162/neco.1997.9.8.1735}.

\bibitem[Johansson et~al.(2012)Johansson, Nation, and Nori]{johansson2012qutip}
J~Robert Johansson, Paul~D Nation, and Franco Nori.
\newblock Qutip: An open-source python framework for the dynamics of open
  quantum systems.
\newblock \emph{Computer Physics Communications}, 183\penalty0 (8):\penalty0
  1760--1772, 2012.
\newblock URL \url{https://doi.org/10.48550/arXiv.1110.0573}.

\bibitem[Kinoshita et~al.(2006)Kinoshita, Wenger, and
  Weiss]{kinoshita2006quantum}
Toshiya Kinoshita, Trevor Wenger, and David~S Weiss.
\newblock A quantum newton's cradle.
\newblock \emph{Nature}, 440\penalty0 (7086):\penalty0 900--903, 2006.
\newblock URL \url{https://doi.org/10.1038/nature04693}.

\bibitem[LeCun et~al.(2015)LeCun, Bengio, and Hinton]{lecun2015deep}
Yann LeCun, Yoshua Bengio, and Geoffrey Hinton.
\newblock Deep learning.
\newblock \emph{nature}, 521\penalty0 (7553):\penalty0 436--444, 2015.
\newblock URL \url{https://doi.org/10.1038/nature14539}.

\bibitem[Liu et~al.(2019)Liu, Li, Sun, and Yu]{liu2019advances}
Yang Liu, Jingfa Li, Shuyu Sun, and Bo~Yu.
\newblock Advances in gaussian random field generation: a review.
\newblock \emph{Computational Geosciences}, pages 1--37, 2019.
\newblock URL \url{https://doi.org/10.1007/s10596-019-09867-y}.

\bibitem[L{\'o}pez-Guti{\'e}rrez and Mendl(2019)]{lopez2019real}
Irene L{\'o}pez-Guti{\'e}rrez and Christian~B Mendl.
\newblock Real time evolution with neural-network quantum states.
\newblock \emph{arXiv preprint arXiv:1912.08831}, 2019.
\newblock URL \url{https://doi.org/10.22331/q-2022-01-20-627}.

\bibitem[Mandel and Wolf(1995)]{mandel1995optical}
Leonard Mandel and Emil Wolf.
\newblock \emph{Optical coherence and quantum optics}.
\newblock Cambridge university press, 1995.
\newblock URL \url{https://doi.org/10.1017/CBO9781139644105}.

\bibitem[Martinez et~al.(2016)Martinez, Muschik, Schindler, Nigg, Erhard, Heyl,
  Hauke, Dalmonte, Monz, Zoller, et~al.]{martinez2016real}
Esteban~A Martinez, Christine~A Muschik, Philipp Schindler, Daniel Nigg,
  Alexander Erhard, Markus Heyl, Philipp Hauke, Marcello Dalmonte, Thomas Monz,
  Peter Zoller, et~al.
\newblock Real-time dynamics of lattice gauge theories with a few-qubit quantum
  computer.
\newblock \emph{Nature}, 534\penalty0 (7608):\penalty0 516--519, 2016.
\newblock URL \url{https://doi.org/10.1038/nature18318}.

\bibitem[Mbeng et~al.(2020)Mbeng, Russomanno, and Santoro]{mbeng2020quantum}
Glen~Bigan Mbeng, Angelo Russomanno, and Giuseppe~E Santoro.
\newblock The quantum ising chain for beginners.
\newblock \emph{arXiv preprint arXiv:2009.09208}, 2020.
\newblock URL \url{https://doi.org/10.48550/arXiv.2009.09208}.

\bibitem[Medina and Semi\~ao(2019)]{PhysRevA.100.012103}
I.~Medina and F.~L. Semi\~ao.
\newblock Pulse engineering for population control under dephasing and
  dissipation.
\newblock \emph{Phys. Rev. A}, 100:\penalty0 012103, Jul 2019.
\newblock \doi{10.1103/PhysRevA.100.012103}.
\newblock URL \url{https://link.aps.org/doi/10.1103/PhysRevA.100.012103}.

\bibitem[Meinert et~al.(2013)Meinert, Mark, Kirilov, Lauber, Weinmann, Daley,
  and N\"agerl]{PhysRevLett.111.053003}
F.~Meinert, M.~J. Mark, E.~Kirilov, K.~Lauber, P.~Weinmann, A.~J. Daley, and
  H.-C. N\"agerl.
\newblock Quantum quench in an atomic one-dimensional ising chain.
\newblock \emph{Phys. Rev. Lett.}, 111:\penalty0 053003, Jul 2013.
\newblock \doi{10.1103/PhysRevLett.111.053003}.
\newblock URL \url{https://link.aps.org/doi/10.1103/PhysRevLett.111.053003}.

\bibitem[Melnikov et~al.(2018)Melnikov, Nautrup, Krenn, Dunjko, Tiersch,
  Zeilinger, and Briegel]{melnikov2018active}
Alexey~A Melnikov, Hendrik~Poulsen Nautrup, Mario Krenn, Vedran Dunjko, Markus
  Tiersch, Anton Zeilinger, and Hans~J Briegel.
\newblock Active learning machine learns to create new quantum experiments.
\newblock \emph{Proceedings of the National Academy of Sciences}, 115\penalty0
  (6):\penalty0 1221--1226, 2018.
\newblock URL \url{https://doi.org/10.1073/pnas.1714936115}.

\bibitem[Mills et~al.(2020)Mills, Ronagh, and Tamblyn]{mills2020controlled}
Kyle Mills, Pooya Ronagh, and Isaac Tamblyn.
\newblock Finding the ground state of spin hamiltonians with reinforcement
  learning.
\newblock \emph{Nature Machine Intelligence}, 2\penalty0 (9):\penalty0
  509--517, 2020.
\newblock URL \url{https://doi.org/10.1038/s42256-020-0226-x}.

\bibitem[Mohseni et~al.(2021)Mohseni, Navarrete-Benlloch, Byrnes, and
  Marquardt]{mohseni2021deep}
Naeimeh Mohseni, Carlos Navarrete-Benlloch, Tim Byrnes, and Florian Marquardt.
\newblock Deep recurrent networks predicting the gap evolution in adiabatic
  quantum computing.
\newblock \emph{arXiv preprint arXiv:2109.08492}, 2021.
\newblock URL \url{https://doi.org/10.48550/arXiv.2109.08492}.

\bibitem[Moon et~al.(2020)Moon, Lennon, Kirkpatrick, van Esbroeck, Camenzind,
  Yu, Vigneau, Zumb{\"u}hl, Briggs, Osborne, et~al.]{moon2020machine}
H~Moon, DT~Lennon, J~Kirkpatrick, NM~van Esbroeck, LC~Camenzind, Liuqi Yu,
  F~Vigneau, DM~Zumb{\"u}hl, G~Andrew~D Briggs, MA~Osborne, et~al.
\newblock Machine learning enables completely automatic tuning of a quantum
  device faster than human experts.
\newblock \emph{Nature communications}, 11\penalty0 (1):\penalty0 1--10, 2020.
\newblock URL \url{https://doi.org/10.1038/s41467-020-17835-9}.

\bibitem[Nielsen(2015)]{nielsen2015neural}
Michael~A Nielsen.
\newblock \emph{Neural networks and deep learning}, volume 2018.
\newblock Determination press San Francisco, CA, 2015.

\bibitem[Or{\'u}s(2014)]{orus2014practical}
Rom{\'a}n Or{\'u}s.
\newblock A practical introduction to tensor networks: Matrix product states
  and projected entangled pair states.
\newblock \emph{Annals of Physics}, 349:\penalty0 117--158, 2014.
\newblock URL \url{https://doi.org/10.1016/j.aop.2014.06.013}.

\bibitem[Peano et~al.(2021)Peano, Sapper, and Marquardt]{peano2019rapid}
Vittorio Peano, Florian Sapper, and Florian Marquardt.
\newblock Rapid exploration of topological band structures using deep learning.
\newblock \emph{Physical Review X}, 11\penalty0 (2):\penalty0 021052, 2021.
\newblock URL \url{https://doi.org/10.1103/PhysRevX.11.021052}.

\bibitem[Poulin et~al.(2011)Poulin, Qarry, Somma, and
  Verstraete]{PhysRevLett.106.170501}
David Poulin, Angie Qarry, Rolando Somma, and Frank Verstraete.
\newblock Quantum simulation of time-dependent hamiltonians and the convenient
  illusion of hilbert space.
\newblock \emph{Phys. Rev. Lett.}, 106:\penalty0 170501, Apr 2011.
\newblock \doi{10.1103/PhysRevLett.106.170501}.
\newblock URL \url{https://link.aps.org/doi/10.1103/PhysRevLett.106.170501}.

\bibitem[Richerme et~al.(2014)Richerme, Gong, Lee, Senko, Smith, Foss-Feig,
  Michalakis, Gorshkov, and Monroe]{richerme2014non}
Philip Richerme, Zhe-Xuan Gong, Aaron Lee, Crystal Senko, Jacob Smith, Michael
  Foss-Feig, Spyridon Michalakis, Alexey~V Gorshkov, and Christopher Monroe.
\newblock Non-local propagation of correlations in quantum systems with
  long-range interactions.
\newblock \emph{Nature}, 511\penalty0 (7508):\penalty0 198--201, 2014.
\newblock URL \url{https://doi.org/10.1038/nature13450}.

\bibitem[Schmitt and Heyl(2020)]{schmitt2019quantum}
Markus Schmitt and Markus Heyl.
\newblock Quantum many-body dynamics in two dimensions with artificial neural
  networks.
\newblock \emph{Phys. Rev. Lett.}, 125:\penalty0 100503, Sep 2020.
\newblock \doi{10.1103/PhysRevLett.125.100503}.
\newblock URL \url{https://link.aps.org/doi/10.1103/PhysRevLett.125.100503}.

\bibitem[Schollw{\"o}ck(2011)]{schollwock2011density}
Ulrich Schollw{\"o}ck.
\newblock The density-matrix renormalization group: a short introduction.
\newblock \emph{Philosophical Transactions of the Royal Society A:
  Mathematical, Physical and Engineering Sciences}, 369\penalty0
  (1946):\penalty0 2643--2661, 2011.
\newblock URL \url{https://doi.org/10.1098/rsta.2010.0382}.

\bibitem[Struck et~al.(2013)Struck, Weinberg, {\"O}lschl{\"a}ger,
  Windpassinger, Simonet, Sengstock, H{\"o}ppner, Hauke, Eckardt, Lewenstein,
  et~al.]{struck2013engineering}
Julian Struck, Malte Weinberg, Christoph {\"O}lschl{\"a}ger, Patrick
  Windpassinger, Juliette Simonet, Klaus Sengstock, Robert H{\"o}ppner, Philipp
  Hauke, Andr{\'e} Eckardt, Maciej Lewenstein, et~al.
\newblock Engineering ising-xy spin-models in a triangular lattice using
  tunable artificial gauge fields.
\newblock \emph{Nature Physics}, 9\penalty0 (11):\penalty0 738--743, 2013.
\newblock URL \url{https://doi.org/10.1038/nphys2750}.

\bibitem[Struck et~al.(2020)Struck, Lindner, Hollmann, Schauer, Schmidbauer,
  Bougeard, and Schreiber]{struck2020robust}
Tom Struck, Javed Lindner, Arne Hollmann, Floyd Schauer, Andreas Schmidbauer,
  Dominique Bougeard, and Lars~R Schreiber.
\newblock Robust and fast post-processing of single-shot spin qubit detection
  events with a neural network.
\newblock \emph{arXiv preprint arXiv:2012.04686}, 2020.
\newblock URL \url{https://doi.org/10.1038/s41598-021-95562-x}.

\bibitem[Sutskever et~al.(2014)Sutskever, Vinyals, and
  Le]{sutskever2014sequence}
Ilya Sutskever, Oriol Vinyals, and Quoc~V Le.
\newblock Sequence to sequence learning with neural networks.
\newblock \emph{Advances in neural information processing systems}, 27, 2014.
\newblock URL
  \url{https://proceedings.neurips.cc/paper/2014/file/a14ac55a4f27472c5d894ec1c3c743d2-Paper.pdf}.

\bibitem[Torlai and Melko(2017)]{PhysRevLett.119.030501}
Giacomo Torlai and Roger~G. Melko.
\newblock Neural decoder for topological codes.
\newblock \emph{Phys. Rev. Lett.}, 119:\penalty0 030501, Jul 2017.
\newblock \doi{10.1103/PhysRevLett.119.030501}.
\newblock URL \url{https://link.aps.org/doi/10.1103/PhysRevLett.119.030501}.

\bibitem[Torlai et~al.(2018)Torlai, Mazzola, Carrasquilla, Troyer, Melko, and
  Carleo]{torlai2018neural}
Giacomo Torlai, Guglielmo Mazzola, Juan Carrasquilla, Matthias Troyer, Roger
  Melko, and Giuseppe Carleo.
\newblock Neural-network quantum state tomography.
\newblock \emph{Nature Physics}, 14\penalty0 (5):\penalty0 447--450, 2018.
\newblock URL \url{https://doi.org/10.1038/s41567-018-0048-5}.

\bibitem[Van~Nieuwenburg et~al.(2017)Van~Nieuwenburg, Liu, and
  Huber]{van2017learning}
Evert~PL Van~Nieuwenburg, Ye-Hua Liu, and Sebastian~D Huber.
\newblock Learning phase transitions by confusion.
\newblock \emph{Nature Physics}, 13\penalty0 (5):\penalty0 435--439, 2017.
\newblock URL \url{https://doi.org/10.1038/nphys4037}.

\bibitem[Vandersypen and Chuang(2005)]{RevModPhys.76.1037}
L.~M.~K. Vandersypen and I.~L. Chuang.
\newblock Nmr techniques for quantum control and computation.
\newblock \emph{Rev. Mod. Phys.}, 76:\penalty0 1037--1069, Jan 2005.
\newblock \doi{10.1103/RevModPhys.76.1037}.
\newblock URL \url{https://link.aps.org/doi/10.1103/RevModPhys.76.1037}.

\bibitem[Wang(2016)]{PhysRevB.94.195105}
Lei Wang.
\newblock Discovering phase transitions with unsupervised learning.
\newblock \emph{Phys. Rev. B}, 94:\penalty0 195105, Nov 2016.
\newblock \doi{10.1103/PhysRevB.94.195105}.
\newblock URL \url{https://link.aps.org/doi/10.1103/PhysRevB.94.195105}.

\bibitem[Wetzel(2017)]{wetzel2017unsupervised}
Sebastian~J Wetzel.
\newblock Unsupervised learning of phase transitions: From principal component
  analysis to variational autoencoders.
\newblock \emph{Physical Review E}, 96\penalty0 (2):\penalty0 022140, 2017.
\newblock URL \url{https://doi.org/10.1103/PhysRevE.96.022140}.

\bibitem[Yang et~al.(2020)Yang, Leng, Yu, Patel, Hu, and
  Pu]{PhysRevResearch.2.012039}
Li~Yang, Zhaoqi Leng, Guangyuan Yu, Ankit Patel, Wen-Jun Hu, and Han Pu.
\newblock Deep learning-enhanced variational monte carlo method for quantum
  many-body physics.
\newblock \emph{Phys. Rev. Research}, 2:\penalty0 012039, Feb 2020.
\newblock \doi{10.1103/PhysRevResearch.2.012039}.
\newblock URL \url{https://link.aps.org/doi/10.1103/PhysRevResearch.2.012039}.

\end{thebibliography}
\providecommand{\noopsort}[1]{}\providecommand{\singleletter}[1]{#1}%

\end{document}